\documentclass[aps,onecolumn,preprint,superscriptaddress,nofootinbib,floats,tightenlines,longbibliography,12pt]{revtex4-1}
\usepackage{amsmath,amssymb,color,mathrsfs,graphicx,verbatim,epsfig, bbm, wasysym, axodraw}
\usepackage[hyperfootnotes=false]{hyperref}
\usepackage{slashed}
\usepackage{cancel}
\usepackage{xcolor}
\allowdisplaybreaks

\usepackage{slashed} 
\usepackage{color, multirow}

\newcommand{\mc}{\mathcal}
\newcommand{\mr}{\mathrm}
\newcommand{\as}{\alpha_s}

\newcommand{\fms}[1]{{#1}\!\!\!/}

\newcommand{\n}{\overline{n}}

\newcommand{\mM}{\mathcal{M}}
\newcommand{\mO}{\mathcal{O}}

\newcommand{\mC}{\mathcal{C}}

\newcommand{\mL}{\mathcal{L}}
\newcommand{\mT}{\mathcal{T}}

\newcommand{\nnn}{\frac{\fms{n}}{2}} 

\newcommand{\eps}{\epsilon}
\newcommand{\e}{\epsilon}

\newcommand{\rc}{r_{\chi}}

\def\be{\begin{equation}}
\def\ee{\end{equation}}
\newcommand{\ba}{\begin{array}}
\newcommand{\ea}{\end{array}}
\newcommand{\bea}{\begin{eqnarray}}
\newcommand{\eea}{\end{eqnarray} }
\newcommand{\bal}{\begin{align}}
\newcommand{\eal}{\end{align}}
\def\bi{\begin{itemize}}
\def\ei{\end{itemize}}
\def\ben{\begin{enumerate}}
\def\een{\end{enumerate}}
\def\beq{\begin{equation}}
\def\eeq{\end{equation}}
\def\bc{\begin{center}}
\def\ec{\end{center}}
\def\bt{\begin{table}}
\def\et{\end{table}}
\def\btb{\begin{tabular}}
\def\etb{\end{tabular}}

\newcommand{\ignore}[1]{}
\renewcommand{\bf}{\textbf}

\newcommand{\GeV}{{\rm\ GeV}}

\newcommand{\TeV}{{\rm\ TeV}}

\newcommand{\amc}{{\sc MadGraph5\textunderscore}a{\sc MC@NLO}}

\DeclareMathOperator{\Li}{Li}

\begin{document}

\baselineskip=18pt


\thispagestyle{empty}
\vspace{20pt}
\font\cmss=cmss10 \font\cmsss=cmss10 at 7pt

\begin{flushright}
\small 
\end{flushright}

\hfill
\vspace{20pt}

\begin{center}
{\Large \textbf
{
End-point resummation in squark decays
}}
\end{center}

\vspace{15pt}
\begin{center}
{ Chul Kim$^{\, a}$, Jeong Han Kim$^{\, b, \, c}$, Seung J. Lee$^{\, d}$ and Jure Zupan$^{\, e}$}
\vspace{30pt}

$^{a}$ {\small \it Institute of Convergence Fundamental Studies and School of Liberal Arts, \\ 
Seoul National University of Science and Technology, Seoul 01811, Korea}

\vskip 3pt
$^{b}$ {\small \it Department of Physics, University of Notre Dame, Notre Dame, IN, 46556, USA}

\vskip 3pt
$^{c}$ {\small \it Department of Physics and Astronomy, University of Kansas, Lawrence, KS, 66045, USA}

\vskip 3pt
$^{d}$ {\small \it Department of Physics, Korea University, Seoul 136-713, Korea and \\
School of Physics, Korea Institute for Advanced Study, Seoul 130-722, Korea}

\vskip 3pt
$^{e}$ {\small \it Department of Physics, University of Cincinnati, Cincinnati, Ohio 45221,USA}

\end{center}

\vspace{20pt}
\begin{center}
\textbf{Abstract}
\end{center}
\vspace{5pt} {\small
We study soft and collinear gluon emission in squark decays to quark--neutralino pair, at next-to-next-to-leading logarithmic (NNLL) accuracy in the end-point region, using Soft Collinear Effective Theory (SCET), and at next-to-leading (NLO) fixed order in the rest of the phase space. As a phenomenological case study we discuss the impact of radiative corrections on the simultaneous measurements of squark and neutralino masses at a linear $e^{+}e^{-}$ collider based on $\sqrt{s} = 3$ TeV Compact Linear Collider (CLIC), and show the softening of distributions in the sum of energies of the first two hardest jets or in the $M_C$ variable. Since the majority of mass measurement techniques are based on edges in kinematic distributions, and these change appreciably when there is additional QCD radiation in the final state, the knowledge of higher-order QCD effects is required for precise mass determinations. 
}

\vfill\eject
\noindent

\newpage


%
\section{Introduction} \label{sec:Intro}

The discovery of Dark Matter (DM) in a collider experiment crucially depends on the ability to measure precisely its properties -- its mass and couplings to visible matter. These are the necessary ingredients to test the hypothesis of a ``WIMP" miracle \cite{Jungman:1995df,Steigman:1984ac, Lee:1977ua, Bertone:2004pz}. Given the importance of such a discovery a number of methods to measure DM mass have been developed \cite{Barr:2010zj,Barr:2011xt, Burns:2008va, Konar:2009qr, Konar:2009wn, Cho:2007qv, Cho:2007dh, Kim:2017awi, Debnath:2017ktz, Bae:2017ebe, Kawagoe:2004rz, Nojiri:2007pq, Cheng:2008mg,Cho:2008tj}.
%
In this paper we are interested in understanding how QCD radiations modifies the precise determination of DM mass. 
Many of the methods for DM mass measurements were developed with 
low energy supersymmetry (SUSY) in mind \cite{Martin:1997ns}. We will thus also use SUSY as an example, though our results do apply more generally. 
 
A significant effort was devoted in measuring DM mass at hadronic colliders. An ingenious method 
was put forward in \cite{Cho:2007qv, Cho:2007dh}, where it was applied to $\tilde g\to q\bar q \chi$ decays in gluino pair production. The mass of $\tilde g$ and $\chi$ can both be measured simultaneously from $m_{T2}$, by computing for each event the value of $m_{T2}$ as a function of an assumed $\chi$ mass, $m_{T2}(m_{\rm trial})$. The envelope of $m_{T2}(m_{\rm trial})$ curves exhibits a kink at $m_{\rm trial}=m_\chi$, where $m_{T2}=m_{\tilde g}$. Measuring the kink determines both masses (for the effect of radiative corrections see \cite{Beneke:2016igc}). For two body decays, e.g., for squark decays, $\tilde q\to q \chi$, the kinks in the distributions appear only once initial state radiation is included \cite{Burns:2008va}. This underscores the importance of radiative corrections for DM mass measurement using kinematical distributions.

In this paper we explore a somewhat simpler case -- the squark pair production in $e^+e^-$ collisions. We focus on a two body decay, $\tilde q\to q \chi$, with $q$ a light quark and $\chi$ a neutralino (for earlier work see \cite{Horsky:2008yi,Hollik:2012rc,Hollik:2013xwa}). Emission of a hard gluon converts this to a three body decay, $\tilde q\to q g \chi$, qualitatively changing the kinematical distributions. Hard gluon emissions, on the other hand, are relatively rare, suppressed by small coupling constant, $\alpha_s \lesssim 0.09$ for $m_{\tilde q}\gtrsim 1~\mr{TeV}$. Most commonly the radiated gluons are either soft or collinear with the outgoing quark, affecting the kinematical distributions in the end-point region where the decay is almost two-body. 
Parameterizing the neutralino energy in the squark rest frame as 
\be
\label{neue}
E_{\chi} = \frac{z M}{2}+ \frac{m_{\chi}^2}{2zM}\ , 
\ee
the end-point region is given by $z\sim 1$. Here $M~(m_{\chi})$ is the squark (neutralino) mass, while the dimensionless variable $z$ takes values,  $z\in [m_{\chi}/M,1]$. Near the end-point the neutralino is maximally boosted and $z$ becomes close to 1. 

The collinear and soft singularities of QCD contributions in the end-point regions lead to large logarithms, $L\sim \ln(1-z)$, in the calculation of the differential decay width, $d\Gamma/dz$. Working to next-to-leading order (NLO) in $\alpha_s$, i.e., to ${\mathcal O}(\alpha_s)$, the Sudakov effects result in large double logarithmic contributions of the form $\as L^2$.
In order to obtain reliable predictions, these logarithms need to be resummed to all orders in $\alpha_s$. At next-to-next-to-leading logarithmic (NNLL) accuracy the resummed decay width is given by 
\be
\label{resumil}
\ln \frac{d\Gamma}{dz} = L f_0(\as L) + f_1(\as L) + \as f_2(\as L),
\ee
with $f_i(\ldots)$ dimensionless functions that are ${\mathcal O}(1)$, counting the large logarithms as $L\sim 1/\alpha_s$. This shows explicitly the dominance of the end-point region, where the first term on the right-hand side (r.h.s.) is the leading contribution. Keeping just the first term would give the result for decay width at leading logarithmic (LL) accuracy, obtained by resumming the double logarithms in the perturbative expansion of the form $\exp(L f_0(\as L) ) = \sum_{k=0} a_k (\as L^2)^{k}$. The second and the third terms on the r.h.s. in Eq. \eqref{resumil}, of NLL and NNLL accuracy,  then resum terms that are additionally suppressed by $\alpha_s$ and $\alpha_s^2$, respectively.

To resum the end-point logarithms we employ soft-collinear effective theory (SCET)~\cite{Bauer:2000ew,Bauer:2000yr,Bauer:2001yt}, which properly describes collinear and soft gluon radiation in the end-point region. The squark decay near the end-point is governed by three distinct scales: hard ($\mu_H$), jet ($\mu_J$), and soft ($\mu_S$) scales.  For large mass splittings, $M-m_{\chi} \sim \mc{O}(M)$, the hard scale $\mu_H$ can be identified with $\mu_H\sim M$. 
The light quark together with radiated collinear gluons forms a collimated jet, controlled by a typical scale $\mu_J\sim M\sqrt{1-z}$. Finally, the soft gluon radiations arise at the scale $\mu_S \sim M(1-z)$. 
Note that the kinematics of this problem is very similar to the decay $B\to X_s \gamma$ in the end-point region, i.e., in the part of the phase space  where the final state photon is close to maximally boosted. The effects of strong  interactions are in this case described by collinear and soft gluon radiations. The factorization formalism for $B\to X_s \gamma$ near the end-point was established in Refs.~\cite{Bauer:2000ew, Bauer:2001yt, Bosch:2004th}.

Similarly to $B\to X_s \gamma$, the differential decay width for $\tilde q\to q \chi$ can be schematically factorized as 
\be 
\frac{d\Gamma}{dz} = H(M,\mu_F) J (M\sqrt{1-z},\mu_F) \otimes S(M(1-z),\mu_F),
\ee
where $H$, $J$, and $S$ are the hard, jet, and collinear functions, respectively. The
`$\otimes$' denotes the  appropriate convolution over $1-z$, while $\mu_F$ is the factorization scale. 
The decay width is independent of the factorization scale, which means that $\mu_F$ can be chosen arbitrarily. 
In general there will be large hierarchies between $\mu_F$ and $\mu_{H,J,S}$, so that one needs to perform renormalization group (RG) evolution for each of the $H$, $J$ and $S$ functions. These RG evolutions in SCET automatically resum the large end-point logarithms. 

Away from the end-point region, where $1-z \sim \mc{O}(1)$, the differential rate is dominated by hard gluon emissions from squark and quark lines, giving the event rate that is ${\mathcal O}(\alpha_s)$. This is of the same order as the NNLL corrections in the end-point region and thus needs to be kept in our expressions.  
We compute these contributions using fixed order calculation at NLO in $\as$.  We smoothly connect the two expressions, valid in the end-point region and away from the end-point regions, giving our final result for the decay width distribution at NNLL+NLO accuracy. 
We use the obtained expressions to perform a numerical study of the impact of QCD corrections in $e^+e^-\to \tilde q \tilde q^*$ events, using a weighted Monte-Carlo simulation.

To compare directly with the experiment our results for the decay widths will still need to be supplemented with a resummation of soft and Coulomb gluon radiation contributions connecting the two squarks, see Refs. \cite{Kim:2014yaa,Beneke:2016kvz} for LHC. These are especially important for slowly moving squarks, i.e.,  at threshold productions, and can even lead to squark bound states \cite{Drees:1993uw,Martin:2008sv,Kim:2014yaa,Batell:2015zla}.

The paper is structured as follows. In Section~\ref{Sec:EFT}, we introduce the necessary ingredients of the effective field theory (EFT) approach to the problem, that includes SCET and heavy Scalar Effective Theory (HSET). The HSET describes soft fluctuations of the heavy squark arising from soft gluon radiations. The HSET and SCET are then used to derive the factorization theorem for the squark decay rate near end-point in Section \ref{sec:endpoint}. The NLO calculation of the decay width in the full kinematical range of $z$ is obtained in Section \ref{FullRange}. Using our results that combine the resummed and fixed calculations, giving the NNLL+NLO accuracy, we perform in Section \ref{sec:pheno} a phenomenological study of squark pair production in $e^+e^-$ annihilation, and then conclude in Section \ref{sec:Summary}. Appendix \ref{AB} contains technical details on $\Delta$-distribution which has been used to regularize infrared (IR) divergences in the fixed NLO calculation.

\section{Construction of Effective Theory Lagrangian}
\label{Sec:EFT}

We are interested in the squark decay, $\tilde q\to q \chi$, where $\chi$ is the dark matter (DM) particle, and how this is affected by QCD radiation. Near the end-point, $\chi$ and a collimated jet are almost back-to-back in the squark rest frame. DM, $\chi$, escapes detection and manifests itself in the detector as missing energy. The quark interacts strongly -- it radiates collinear gluons and quark-antiquark pairs, which form a collimated jet. In addition,  there is soft gluon radiation in the event, which does not have a preferred direction. 

As explained in the Introduction, the decay is governed by three distinct scales, $\mu_H$, $\mu_J$, and $\mu_S$. We use EFTs to deal with the  hierarchies between the three scales and the associated large logarithms. We first integrate out the hard interactions, where the relevant hard scale, $\mu_H$, is comparable to the squark mass $M$. At energy scales below $\mu_H$ we then have only collinear and soft degrees of freedom. The light quark and the collinear gluon describe collinear interactions for the collimated jet. 
Also the soft mode decoupled from the collinear quark and the heavy squark describes soft gluon radiations near the end-point. 
SCET is the appropriate EFT that describes collinear and soft modes and their interactions. It provides a systematic way to decouple soft modes from the collinear field. This is very useful when proving factorization in the end-point region. The interactions of heavy squark are described by the HSET, which is obtained by integrating out the hard gluon modes and the squark mass $M$.
In the rest of this section we show how SCET and HSET are constructed. The decay rate of the heavy squark is calculated in the subsequent section.  

\subsection{Decay Lagrangian at the hard scale}
We take $\chi$ to be a Majorana fermion. This is the case in the MSSM where $\chi$ is the lightest supersymmetric particle (LSP) -- assumed to be the lightest neutralino. The most general Lagrangian describing a two-body decay of a color triplet scalar, $\tilde q$, to a quark, $q$, and a Majorana fermion $\chi$, is given by
\footnote{For a Dirac fermion $\chi$ there are two additional terms in \eqref{EFT1}, $B_L'(\mu)\big(\overline{q}_L P_R \chi^c \big)\tilde{q}$ and $ B_R'(\mu)\big(\overline{q}_R P_L \chi^c \big) \tilde{q} $. }
\begin{equation}
\label{EFT1} 
\mL_{\rm{int}} = \sum_{L,R} B_i (\mu)O_i(\mu) + \mathrm{h.c.}  
= B_L(\mu)\,\big(\overline{q}_L P_R \chi \big)\tilde{q} + B_R(\mu)\,\big(\overline{q}_R P_L \chi \big) \tilde{q} + \mathrm{h.c.}, 
\end{equation} 
where we are using the four-component notation with $P_{L,R} = (1\mp \gamma_5)/2$. The dimensionless Wilson coefficients $B_{L,R}$ encode the new physics as well as strong interactions above the hard scale $\mu_H\sim M$. Our analysis applies to MSSM, but is also more general and applies to any decays of the form $\tilde q\to q \chi$, where $\tilde q$ is a color triplet scalar.

In the MSSM for each quark flavor there are two squarks, $\tilde q_{1,2}$, so that the above Lagrangian modifies to 
\begin{equation}
\label{EFT-MSSM} 
\mL_{\rm{int}} =\sum_{i=1,2}B_{Li}(\mu)\,\big(\overline{q}_L P_R \chi \big)\tilde{q_i} + B_{Ri}(\mu)\,\big(\overline{q}_R P_L \chi \big) \tilde{q_i} + \mathrm{h.c.}.
\end{equation} 
The tree level expressions for the Wilson coefficients are, neglecting flavor violating effects,
\beq
B_{Li}=C_{LL} L_{\tilde q_i}+C_{LR} R_{\tilde q_i},\qquad B_{Ri}=C_{RL} L_{\tilde q_i}+C_{RR} R_{\tilde q_i},
\eeq
with
\bea
C_{LL}&=&-\sqrt{2}\big[g T_3^q N_{12}+g' (Q_q-T_3^q)N_{11}\big], \qquad C_{RR}=\sqrt{2}g'Q_qN_{11}^*, \label{C_LL}\\
C_{RL}&=&C_{LR}^*=- \sqrt2 m_q \big(N_{14}^* \delta_{qu} +N_{13}^*  \delta_{qd}\big)/v, \label{C_RL}
\eea
with $g,g'$ the weak and hypercharge gauge couplings, $Q_q$ the electric charge of quark $q$, and $T_3^q$ the weak isospin, while $R_{\tilde q_1}=L_{\tilde q_2}^*=\cos (\theta_{\tilde q})$, and $L_{\tilde q_1}=-R_{\tilde q_2}^*=\sin(\theta_{\tilde q})$, with $\theta_{\tilde q}$ the mixing angle rotating the squark gauge eigenstates $\tilde q_{R,L}$  to mass eigenstates $\tilde q_{1,2}$. The $\tilde q_R$--$\tilde q_L$ mixing is usually important only for the third generation squarks, while for the first two generations gauge and mass eigenstates coincide, $\theta_{\tilde q}=0$. The neutralino mixing matrix is denoted by $N_{ij}$. If LSP is mostly gaugino then $N_{11,12}\gg N_{13,14}$ and thus $\tilde q_L\to q_L\chi$ and $\tilde q_R\to q_R \chi$ for the first two generations. For well-tempered neutralino, on the other hand, all terms in  \eqref{C_LL},\eqref{C_RL}  may be important. 

\subsection{EFTs for the end-point region}
We restrict ourselves to the case where quark mass can be neglected compared to $M$. We will work in the squark rest frame, so that its four-velocity $v^{\mu}$ is given by $v^{\mu} = (1,{\bf 0})$. We orient the coordinate system such that jet goes in the $z$ direction, i.e., that, neglecting its mass, it is on the light cone $n^\mu=(1,0,0,1)$. We also introduce the opposite light cone four-vector $\bar n^\mu=(1,0,0,-1)$, so that $n^2 =\n^2 =0$,  $n\cdot\n = 2$ and $p^{\mu}_{\tilde q} = M v^{\mu} = {M} (n^{\mu} +\n^{\mu})/2$.  We will use light-cone coordinates, in which  a four-momentum $p^\mu$ is  given by $p^\mu=(\bar n\cdot p, p_{\perp},n\cdot p)$.

The effective field theory to reproduce low energy physics in full QCD is obtained by integrating hard degrees of freedom. For instance, the hard gluon exchanges between the heavy squark and the light quark are integrated out. The Wilson coefficients $B_{L,R}$ in Eq.~(\ref{EFT1}) thus get modified to $\mC_{L,R} (\mu)$ (see Eq.~\eqref{EFTlow} below). The resultant EFT is valid at the scale $\mu < \mu_H\sim M$. And the remaining degrees of freedom in EFT are collinear and soft fields scaling as $p_{c} = M(1,\lambda,\lambda^2)$ and $p_s = M(\lambda^2,\lambda^2,\lambda^2)$ respectively. Here $\lambda$ is a small expansion parameter in EFT. For the squark decay near end-point, $\lambda$ is given as $\sim \sqrt{1-z}$. 

In the heavy squark sector, after integrating out hard fluctuations as well as heavy squark mass $M$, the heavy squark only interacts with soft gluons. Then full QCD Lagrangian for the heavy squark can be matched onto HSET Lagrangian,
\begin{equation}
\label{HSET}
\mL_{\mathrm{HSET}} = \phi_v^{*} v\cdot iD_s \phi_v - \frac{1}{2M} \phi_v^* D_s^2 \phi_v+{\cal O}(1/M^2),
\end{equation}
where $\phi_v$ is the squark field in HSET,
\begin{equation}
\label{hsq}
\tilde{q} (x) = \frac{1}{\sqrt{2M}} e^{-iMv\cdot x} \phi_v (x),
\end{equation}
The covariant derivative $D_s^{\mu} = \partial^{\mu} - ig A_s^{\mu,a} T^a$ includes only the soft gluon field. 
The second term in \eqref{HSET} is $\mO(1/M)$ with a coefficient that is fixed by reparametrization invariance. We work at leading order in $1/M$ expansion, and thus only keep the first term in \eqref{HSET}.

The light quark field matches onto $n$-collinear field in SCET so that 
\beq
q(x) = \sum_{\tilde{p}} e^{-i\tilde{p}\cdot x}  q_{n,p} (x) = \sum_{\tilde{p}} e^{-i\tilde{p}\cdot x} \Bigl(\xi_{n,p} (x) + \xi_{\n,p} (x) \Bigr), 
\eeq
where
\beq
\xi_{n,p} (x)=\frac{\fms{n}\fms{\n}}{4}q_{n,p}(x), \quad  \xi_{\n,p} (x)=\frac{\fms{\n}\fms{n}}{4}q_{n,p}(x),
\eeq
and thus $\fms n\xi_{n,p} =\fms \n \xi_{\n,p} =0$. The summation is over large label momenta given by $\tilde{p}^{\mu} = \n\cdot {p} n^{\mu}/2 + {p}_{\perp}^{\mu}$ that differ by soft fluctuations. The field  $\xi_{\n}$ is suppressed by $\lambda $, and is thus not present as an external quark field in our analysis of squark decays since we work to leading order (LO) in the $1/M$ expansion. Integrating out $\xi_{\n}$ the collinear interactions can be expressed entirely in terms of $\xi_n$. The resulting LO SCET Lagrangian for collinear fields can be found in, e.g., Ref. \cite{Bauer:2000yr}.

The decay Lagrangian \eqref{EFT1} matches onto the HSET+SCET effective decay Lagrangian, appropriate for describing the squark decays in the end-point region,
\begin{equation} 
\begin{split}
\label{EFTlow} 
\mL_{\rm{int}} ^{\rm eff} &= \Big[\sum_{\tilde p} \frac{ \mC_L (\mu)}{\sqrt{2M}} e^{-i(Mv-\tilde{p})\cdot x}
\big(\bar{\xi}_{n,p} W_n P_{R}\chi \big) \phi_{v} (x) + \mathrm{h.c.}\Big]+ \big[L \leftrightarrow R\big] 
\\
&\equiv \sum_{\tilde p}\Bigl[e^{-i(Mv-\tilde{p})}\mC_L(\mu) \mO_L (x,\mu) + \mathrm{h.c.}\Bigr] + \sum_{\tilde p}\Bigl[e^{-i(Mv-\tilde{p})}\mC_R(\mu) \mO_R (x,\mu) + \mathrm{h.c.}\Bigr], 
\end{split}
\end{equation} 
In the sum only the $\tilde{p}$ that satisfy momentum conservation are selected. The hard gluon exchanges are encoded in Wilson coefficients $\mC_{L,R}$ (obtained from $B_{L,R}$ in \eqref{EFT1}), while collinear gluons emitted from the heavy squark yield the collinear Wilson line
 \begin{equation}
W_n (x) = \mathrm{P} \exp \Biggl(ig \int^x_{-\infty} ds \n\cdot A_n^a (s\n^{\mu}) T^a\Biggl).
\label{CW}
\end{equation}
Here $A_n^{\mu}$ is $n$-collinear gluon field and `P' indicates the path-ordered integral.
 
To show the factorization of soft and collinear interactions it is useful to perform field redefinitions,   $\xi_n \to Y_n \xi_n$, $A_n^{\mu} \to Y_n A_n^{\mu} Y_n^{\dagger}$, and $\phi_v \to Y_v \phi_v$~\cite{Bauer:2001yt}, factoring out the soft Wilson lines in the $n$ and $v$ directions,  $Y_{n,v}$, 
\begin{equation}
Y_{\mathrm{v}} (x) = \mathrm{P} \exp \Biggl(ig \int^x_{-\infty} ds \mathrm{v}\cdot A_s^a (s\mathrm{v}^{\mu}) T^a\Biggl),~~~\mathrm{v}^{\mu} = n^{\mu}, v^{\mu}.
\label{SW}
\end{equation}
The path of integration over $s\in [-\infty,x]$  indicates that the dressed collinear or squark field is incoming. For the outgoing particles  the integration path is over $s\in [x,+\infty]$, giving for the soft Wilson lines~\cite{Chay:2004zn},
\begin{equation}
\tilde{Y}^{\dagger}_{\mathrm{v}} (x) = \mathrm{P} \exp \Biggl(ig \int_x^{+\infty} ds \mathrm{v}\cdot A_s^a (s\mathrm{v}^{\mu}) T^a\Biggl),~~~\mathrm{v}^{\mu} = n^{\mu}, v^{\mu}.
\label{SWtd}
\end{equation}
In the LO SCET and HSET Lagrangian the interactions between soft gluons and the redefined collinear fields, $\xi_n$, $A_n^\mu$, and between the soft gluons and the heavy squark field $\phi_v$, drop out (that is, at LO there are no interactions between collinear and soft fields, and no interactions between redefined heavy squark and soft fields). The effects of soft gluons are thus moved into the effective decay Lagrangian, where they appear as a product of two soft Wilson lines in $n$ and $v$ directions, 
\begin{equation}
\label{EFTlredef} 
\begin{split}
\mL_{\rm{int}} ^{\rm eff}= \sum_{i=L,R} \sum_{\tilde p} \mC_i (\mu) e^{-i(Mv-\tilde{p})\cdot x} \mO_i^a \chi_a + \mathrm{h.c.},  
\end{split}
\end{equation} 
with
\beq\label{mO}
 {\cal O}^a_{L,R}(\mu)=\frac{1}{\sqrt{2M}}\big(\bar{\xi}_{n,p} W_n P_{R,L} \big)^a \tilde{Y}_n^{\dagger} Y_v \phi_{v} (x).
 \eeq
In Eq. \eqref{EFTlredef} a summation over Dirac four-component index $a$ is implied.  From now on we will use the form of EFT Lagrangian given in Eq. \eqref{EFTlredef}, i.e., with $\xi_n$, $A_n^\mu$ and $\phi_v$ denoting the redefined fields that do not couple to soft gluons and quarks at LO.

\section{Differential Decay Rate at the end-point}
\label{sec:endpoint}
The total decay rate for $\tilde{q} \to \chi^0 q_L$ averaged over the squark color is 
\begin{equation} 
\label{fullde} 
\Gamma (\tilde{q} \to \chi^0 q_L) =  \frac{1}{2M} \int \frac{d^3 p_{\chi}}{(2\pi)^3}\frac{1}{2E_{\chi}} \mT_L (E_{\chi}, m_{\chi}, M),
\end{equation} 
where $\mT_{L} (E_{\chi}, m_{\chi}, M)$ is related to the matrix elements squared for squark decays into left-handed quarks 
\begin{equation} 
\label{TE}
\mT_{L} (E_{\chi}, m_{\chi}, M) = \sum_X (2\pi)^4 \delta(p-p_{\chi} -p_X) |\mM_{L}|^2. 
\end{equation}  
Explicitly, the matrix elements squared are 
\begin{equation} 
\label{ML}
|\mM_L|^2 = {|B_L (\mu)|^2}\langle \tilde {q} |\tilde{q}^{*\alpha} (P_L q^{\alpha})_a | X \rangle \langle X| (\bar{q}^{\beta} P_R)_b \tilde{q}^{\beta} |\tilde{q}\rangle (\fms{p}_{\chi} - m_{\chi})_{ba}, 
\end{equation} 
with the summation over color indices, $\alpha, \beta$, and Lorentz indices, $a,b$ implied. We do not show color index of squark external state: we always consider color-averaged initial states, hence $\langle \tilde{q} | \cdots |\tilde{q} \rangle = \langle \tilde{q}^{\alpha} | \cdots |\tilde{q}^{\alpha} \rangle /N_c$ in our convention.
Note that in Eq.~(\ref{ML}) we already used the fact that neutralinos are not charged under QCD and thus only contribute as $\sum_{{\rm spin}} u_b \bar u_a= (\fms{p}_{\chi} - m_{\chi})_{ba}$. 
For $\tilde{q}\to \chi^0 q_R$ decays the same results apply, but with $L \leftrightarrow R$. The interference terms between the two decays are $m_q/M$ suppressed and can be safely neglected.

At tree level the decay rate is given by  
\begin{equation} 
\label{tree} 
\Gamma^{(0)} (\tilde{q} \to \chi^0 q_L) = \frac{|B_L|^2}{16\pi} M(1-\rc)^2,
\end{equation} 
with 
\begin{equation} 
\label{eq:rchi}
r_\chi=m_\chi^2/M^2 \;.
\end{equation}

\subsection{Factorization theorem near the end-point}
\label{EndPoint}

We start by reviewing the decay kinematics near the end-point, where in the final states we have an energetic neutralino and a collimated jet as well as soft gluons. Following Eq.~\eqref{neue}, we define the kinematic variable
\begin{equation} 
\label{eq:zval}
z = \frac{n\cdot p_{\chi}}{M},
\end{equation}
in terms of which the neutralino and other final states momenta are given by
\begin{equation} 
\label{pchi}
p_{\chi}^{\mu} = \frac{M}{2} \Big(\frac{r_{\chi}}{z} {n^{\mu}} + z  {\n^{\mu}}\Big),  \quad p_X \equiv p_{J}^{\mu} + p_S^{\mu}= \frac{M}{2} \Big[\Big(1-\frac{r_{\chi}}{z}\Big) {n^{\mu}} + \big(1-z\big)  {\n^{\mu}}\Big].
\end{equation}
Here $z$ can take values $z \in [\sqrt{r_\chi},1]$ (we oriented the coordinate system such that $\vec p_\chi$ is always along negative $z$-axis, while $\vec p_X$ is along positive $z$-axis). The missing energy is $E_\chi=v\cdot p_\chi=(r_\chi/z+z)M/2$, and therefore one can use $z$ and $E_\chi$ interchangeably. 
The invariant mass of the collinear and soft final states is $p_X^2=M^2(1-r_\chi/z)(1-z)$. The limit of a very collimated jet, $p_X^2\to 0$, is thus obtained in the limit $z\to 1$.  

In the end-point region, $1-z\ll 1$, we can apply  the  HSET and SCET formalism, introduced in the previous section. The differential rate is
\begin{equation} 
\label{fullde} 
\frac{d\Gamma (\tilde{q} \to \chi^0 q_L)}{d E_\chi} =  \frac{(E_{\chi}^2-p_{\chi}^2)^{1/2}}{8\pi^2 M} \mT_L (E_{\chi}, m_{\chi}, M),
\end{equation} 
with $
\mT_{L}  = \sum_X (2\pi)^4 \delta(p-p_{\chi} -p_X) |\mM_{L}|^2,  
$ where $|\mM_L|^2$ is calculated in HSET+SCET,
\begin{equation} 
|\mM_L|^2 = 2 M {|\mC_L (\mu)|^2} \langle \phi_v | \mO_{L,a}^\dagger | X \rangle \langle X| \mO_{L,b} |\phi_v\rangle (\fms{p}_{\chi} - m_{\chi})_{ba}  + \mO(1/M). 
\end{equation} 
The EFT operators $\mO_{L,R}$ are given in Eq.~\eqref{mO}, and  $|\tilde{q}\rangle = \sqrt{2M} |{\phi}_{v} \rangle$ up to $1/M$ corrections. 
The Wilson coefficients $\mC_{L,R}$ can be decomposed into 
\beq
\label{eq:CLR:decomp}
\mC_{L,R} = B_{L,R}\times C_{L,R},
\eeq
where $B_{L,R}$ are the unknown new physics Wilson coefficients in Eq.~(\ref{EFT1}) and $C_{L,R}$ are the Wilson coefficients as a result of integrating out the squark mass $M$ and the hard gluon exchanges between squark and quark. The NLO matching onto HSET+SCET has been performed in Section \ref{sec:rad:corr}, with the result for $C_{L,R}$ given in Eq. \eqref{CLR}.

In the remainder of this section we use the operator production expansion (OPE) to arrive at a more practically useful expression for $\mT_{L}$.  We first use the completeness relation, $\sum_X |X\rangle \langle X| =1$, to rewrite $\mT_{L}$ as (neglecting $\mO(1/M)$ corrections)
\beq\label{TE1}
\mT_L = 2 M |\mC_L(M,m_{\chi},\mu)|^2 
\int\negthickspace d^4 y\, e^{i(Mv-p_{\chi}-\tilde{p}_c)\cdot y} (\fms{p}_{\chi} - m_{\chi})_{ba}
 \langle \phi_{v} | \mO_{L,a}^{\dagger} (y) \mO_{L,b} (0) | \phi_{v} \rangle.
\eeq
The large label momentum $\tilde p_c$ is the $n^\mu$-component of $p_X^\mu$, given in Eq.~\eqref{pchi}. The phase in (\ref{TE1}) is therefore equal to
\begin{equation}
\label{phase} 
Mv^{\mu}-p_{\chi}^{\mu}-\tilde{p}_c^{\mu} = M(1-z){\n^{\mu}}/{2},  
\end{equation} 
so that
\begin{equation} 
\begin{split}
\label{TE2}
\mT_L (E_{\chi}, m_{\chi}, M)&=z M^2 |\mC_L(M,m_{\chi},\mu)|^2 
\int d^4 y e^{iM(1-z)y_-/2} 
\\
&\times \langle 0 | (\tilde{Y}_n^{\dagger}Y_v)^{\alpha\beta} (y) (W_n^{\dagger} P_L \xi_n (y))^{\beta}_a 
~(\bar{\xi}_n P_R \fms{\bar n} W_n)_a^{\gamma} (0) (\tilde{Y}_n Y_v^{\dagger})^{\gamma\alpha} (0) | 0 \rangle, 
\end{split}
\end{equation} 
where we used the shorthand notation $y_-=\n\cdot y $ ($y_+=n\cdot y$) and that $\phi_{v}^{\alpha} | \phi_{v}^{\beta} \rangle = \delta^{\alpha\beta}$. 

The collinear field $\xi_n$ describes an inclusive jet in $n$-direction. The corresponding jet function is defined as 
\beq
\begin{split}
\label{jet} 
 \langle 0 | (W_n^{\dagger} \xi_{n,p} (y))^{\alpha}_a (\bar{\xi}_n W_{n,p})_b^{\beta} (0) | 0 \rangle 
&=  \delta^{\alpha\beta} \Bigl(\nnn\Bigr)_{ab}  \int \frac{d^4 k}{(2\pi)^3} e^{-ik\cdot y} J_n (k_+ ; \n \cdot \tilde{p}) 
 \\
&=\delta^{\alpha\beta} \Bigl(\nnn\Bigr)_{ab} \delta(y_+)\delta(y_{\perp})  \int dk_+ e^{-ik_+ y_-/2}J_n (k_+ ; \n \cdot \tilde{p})
  .
\end{split}
\eeq
At LO in $\alpha_s$ the jet function is simply $J_n^{(0)} = \delta(k_+)$. 

The product of Wilson lines in Eq.~\eqref{TE2} forms the soft 
function $S(l_+)$, defined as 
\begin{equation} 
\label{soft} 
S(l_+,\mu) = \mathrm{Tr} \langle 0 | Y_v^{\dagger} \tilde{Y}_n \delta(l_++n\cdot i\partial) \tilde{Y}_n^{\dagger} Y_v |0 \rangle,
\end{equation}  
so that its Fourier transform is 
\begin{equation} 
\label{eq:soft:FF}
\mathrm{Tr} \langle 0 | Y_v^{\dagger} \tilde{Y}_n ({y_-}) \tilde{Y}_n^{\dagger} Y_v (0) |0 \rangle = \int dl_+ e^{-il_+ y_-/2} S(l_+).
\end{equation} 

The expression for $\mT_{L}$ in Eq.~(\ref{TE2}) can therefore be written as the convolution of the soft function, Eq. \eqref{eq:soft:FF}, and the jet function, Eq.~\eqref{jet}, 
\begin{equation} 
\label{TEf} 
\mT_L (z, m_{\chi}, M) = 4 \pi |\mC_L(M,m_{\chi},\mu)|^2 zM^2 \int dl_+ S(l_+,\mu)
J_n \big(M(1-z)-l_+,\mu ; M(1-{r_\chi}/{z})\big) .
\end{equation} 

Collecting all the terms, the differential decay rate $d\Gamma/dz$ in the $z\to 1$ limit  can thus be written as 
\begin{equation} 
\begin{split}
\label{didec}
\frac{d\Gamma}{dz} (\tilde{q} \to \chi^0 X_{q_L}) =&
\frac{M^2}{16\pi} |\mC_L(M,x,\mu)|^2 \Bigl(z-\frac{r_\chi}{z}\Bigr)^2
 \\
&\times
\int dl_+ 
J_n \big(M(1-z)-l_+,\mu ; M(1-{r_\chi}/{z})\big) S(l_+,\mu),
\end{split}
\end{equation} 
where the large momentum, $M(1-r_\chi/z)$, in the jet function can be further simplified to $M(1-r_\chi)$, neglecting $\mO(1-z)$ corrections. 

The factorization formula \eqref{didec}  is similar to the one for $B\to X_s \gamma$~\cite{Bauer:2000ew, Bauer:2001yt, Bosch:2004th}. The main difference are the hard interactions. 
Another difference is that the soft function in the $\tilde q\to \chi q$ decay would be treated perturbatively. For the squark mass $M \gtrsim \mc{O}(\mr{1 TeV})$, the typical soft scale $\mu_S \sim M(1-z)$ would be of a few tens of GeV, and it is much larger than the hadronic scale $\Lambda_{\rm{QCD}}\lesssim 1$ GeV.
Unlike $B\to X_s \gamma$, where the predictions in the end-point region are given in terms of the nonperturbative $B$ meson shape functions, the nonperturbative physics here affects the region of phase space less than $1-z \sim \mO(\Lambda_{\rm{QCD}}/M)\sim 10^{-3}$, which does not significantly change our phenomenological conclusions in Sec. \ref{sec:simulation}.

\subsection{Radiative Corrections}
\label{sec:rad:corr}
We now return to the calculation of $C_{L,R}$ in Eq. \eqref{eq:CLR:decomp}. At scale $\mu \sim M$, we need to integrate out the heavy squark mass $M$ and the hard gluon fluctuations of order $M$, matching onto HSET+SCET. 
For this we match QCD calculations in $\as$ for the effective operators in full QCD (Eq.~(\ref{EFT1})) and ones in SCET+HSET (Eq.~\eqref{EFTlow}). And we obtain the Wilson coefficients $C_{L,R}$ at the higher order in $\as$. At tree level the matching is trivial, resulting in $C_{L,R}=1$. 

\begin{figure}[t]
\begin{center}
\includegraphics[height=5.5cm]{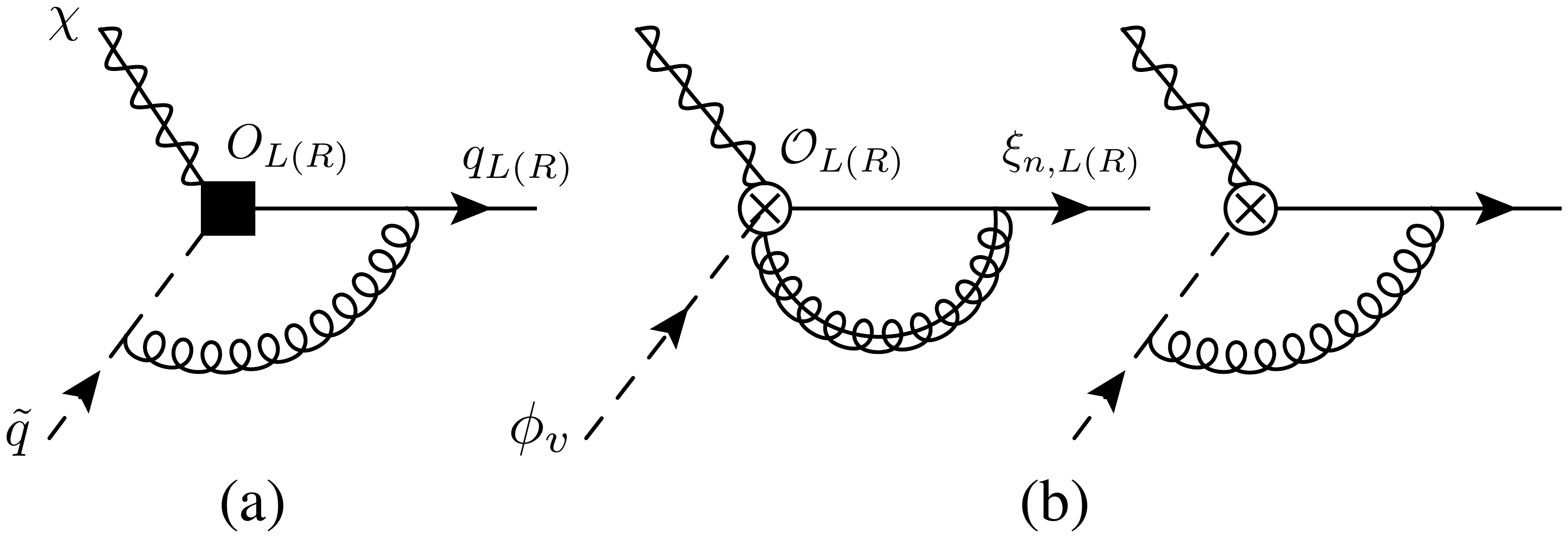}
\end{center}
\vspace{-0.3cm}
\caption{\label{fig1} \baselineskip 3.0ex 
One loop diagrams for matching between full QCD (a) and SCET+HSET (b). Here $O_{L(R)} = \big(\overline{q}_{L(R)} P_{R(L)} \chi \big)\tilde{q}$ and $\mO_{L(R)} = \big(\bar{\xi}_{n,L(R)} W_n P_{R(L)} \chi \big) \phi_{v}$. In diagrams (b), the first (second) diagram is for collinear (soft) gluon exchange. 
Self energy diagrams are also needed for the matching process. 
} 
\end{figure}

For NLO results of $C_{L,R}$ we compute Feynman diagrams shown in Fig.~\ref{fig1} and self energy diagrams at one loop. Throughout this paper, we employ dimensional regularization in the $\overline{\rm MS}$ scheme with $D=4-2\eps$ in order to handle ultraviolet (UV) singularity. As a result we obtain  
\begin{equation} 
\begin{split}
\label{CLR}
C_{L,R} (M,\rc,\mu) =& 1+\frac{\alpha_s}{4\pi}C_F \Biggl[-\Bigl(\frac{9}{4}+\frac{\pi^2}{12}+\ln\frac{\mu^2}{M^2} + \frac{1}{2}\ln^2\frac{\mu^2}{M^2}\Bigr)  
\\
&+2\ln(1-\rc)\Bigl(1+\ln\frac{\mu^2}{M^2}\Bigr) - \ln^2(1-\rc) + 2\Li_2\Bigl(\frac{\rc}{\rc-1}\Bigr)\Biggr]\ .\end{split}
\end{equation}
Here $\Li_2(x)$ is the dilogarithm function. The anomalous dimensions for $C_{L,R}$ are given 
\begin{equation} 
\label{gammaC}
\gamma_{C} = \frac{\mu}{C_{L,R}(\mu)} \frac{d}{d\mu} C_{L,R}(\mu) = -\frac{\alpha_s}{2\pi}C_F \Bigl(1+\ln\frac{\mu^2}{M^2(1-\rc)^2}\Bigr)+\mO(\alpha_s^2).
\end{equation} 

Note that full Wilson coefficients $\mC_{L,R}=B_{L,R}C_{L,R}$ involve the unknown new physics Wilson coefficients $B_{L,R}$. But, since the effective interaction Lagrangian in \eqref{EFTlow} should be scale invariant, we can compute $\gamma_{\mathcal{C}_{L,R}}$ by considering the renormalization behavior of the effective operators $\mO_{L,R}$ in SCET+HSET.  
The relation between the bare and renormalized effective operators can be written as $Z_{\mO}^{L,R} \mO_{L,R}^R = Z_{\phi_v}^{1/2} Z_{\xi}^{1/2} \mO_{L,R}^B$. Here,  
\beq
\label{eq:ZqtildeZq}
Z_{\phi_v} = 1+ \alpha_s C_F/(2\pi\epsilon),\quad~\text{and}~~Z_{\xi} = 1 - \alpha_s C_F/(4\pi\epsilon),
\eeq
are the wave function renormalizations for the heavy squark and the collinear quark fields respectively. 
Computing the one loop diagrams in Fig.~\ref{fig1}-(b) gives 
\begin{equation} 
\label{Zeft}
Z_{\mO}^{L} =  Z_{\mO}^{R} = 1+ \frac{\alpha_s}{4\pi} C_F \Biggl[\frac{1}{\eps^2} + \frac{1}{\eps}\Bigl(\frac{5}{2} + \ln \frac{\mu^2}{(\n\cdot p)^2 (n\cdot v)^2} \Bigr)\Biggr],
\end{equation} 
where $\n\cdot p$ is the large momentum component for the quark and equals $\n\cdot p=M(1-\rc)$ in the $z\to 1$ limit, while $n\cdot v=1$ in the squark rest frame. 
From Eq.~(\ref{Zeft}) we obtain the anomalous dimension for $\mC_{L,R}$ satisfying the RG equation, $d/(d\ln\mu) \mC_{L,R} = \gamma_{\mathcal{C}}~\mC_{L,R}$, as follows 
\begin{equation} 
\label{anomalC} 
\gamma_{\mathcal{C}} = \Bigl(\mu\frac{\partial}{\partial\mu} + \frac{\partial}{\partial g}\Bigr) \ln Z_{\mO}^{L,R}  
= - \frac{\alpha_s}{2\pi} C_F \Bigl(\frac{5}{2} + \ln \frac{\mu^2}{M^2(1-\rc)^2}\Bigr). 
\end{equation} 
Since $\mC_{L,R}=B_{L,R} C_{L,R}$, from Eqs.~(\ref{gammaC}) and (\ref{anomalC}) the anomalous dimensions for the unknown $B_{L,R}$ are obtained as 
\begin{equation} 
\label{gammaB}
\gamma_{B} = \frac{\mu}{B_{L,R}(\mu)} \frac{d}{d\mu} B_{L,R}(\mu) = -3\frac{\alpha_s}{4\pi}C_F +\mO(\alpha_s^2).
\end{equation}

In order to employ the standard plus distribution for the radiative corrections in Eq. (\ref{didec}), it is convenient to introduce dimensionless jet and soft functions,
\begin{align} 
\label{dimljet} 
\bar{J}_n (x,\mu; M(1-\rc)) &= y M J_n (M(1-z)-l_+,\mu;M(1-\rc))\\
\label{dimlsoft}
\bar{S} (y,\mu) &= M S(l_+,\mu),
\end{align}
where $y$ is related to $l_+$ through
\beq
l_+ = M(1-y), \quad\text{~while~} x=z/y.
\eeq
The two new  variables are defined in the interval $z\le x,y \le 1$.  The limit of soft momenta in the soft function corresponds to $y\to 1$. 
In terms of the dimensionless jet and soft functions, the differential decay rate in Eq.~(\ref{didec}) can be rewritten as 
\begin{equation} 
\label{didec1}
\frac{d\Gamma}{dz}(\tilde{q} \to \chi^0 X_{q_L}) =
\frac{M}{16\pi} \Bigl(z-\frac{\rc}{z}\Bigr)^2 |\mC_L(M,\rc,\mu)|^2\int^1_z \frac{dx}{x}
\bar{J}_n \Bigl(x,\mu; M(1-\rc))\Bigr) \bar{S}(z/x,\mu).
\end{equation} 

We computed the jet and the soft functions at  next-to-leading order (NLO) in $\alpha_s$, and the results read ~\cite{Bosch:2004th,Bauer:2003pi}
\begin{align} 
\begin{split}
\label{nlojet}
\bar{J}_n (x,\mu) =& \delta(1-x) + \frac{\alpha_s}{2\pi} C_F \Biggl\{
\delta(1-x)\Biggl[\frac{7}{2}-\frac{\pi^2}{2} + \frac{3}{2} \ln \frac{\mu^2}{Q^2}+\ln^2 \frac{\mu^2}{Q^2}\Biggr] \\
&~~~-\frac{1}{(1-x)_+} \Biggl[2 \ln\frac{\mu^2}{Q^2}+\frac{3}{2} \Biggr]+2\Biggl( \frac{\ln(1-x)}{1-x}\Biggr)_+\Biggr\},
\end{split}
 \\
 \begin{split}
\label{nlosoft}
\bar{S} (y,\mu) =& \delta(1-y) + \frac{\alpha_s}{2\pi} C_F \Biggl\{
\delta(1-y)\Biggl[-\frac{\pi^2}{12} + \ln \frac{\mu^2}{M^2}-\frac{1}{2} \ln^2 \frac{\mu^2}{M^2}\Biggr] \\
&~~~+\frac{2}{(1-y)_+} \Biggl[-1 +\ln\frac{\mu^2}{M^2} \Biggr]-4\Biggl( \frac{\ln(1-y)}{1-y}\Biggr)_+\Biggr\},
\end{split}
\end{align}
where $Q^2 = M^2(1-\rc)$ and $(\ldots)_+$  the standard plus distribution. Note that both of the above results are infrared finite. The logarithms are minimized at $\mu=Q(1-x)^{1/2}$ and $\mu=M(1-y)$ for the jet and soft functions, respectively.  

We can check that the obtained differential decay rate does not depend on the scale choice, $\mu$, to the order we are working. Differentiation with respect to $d(\log\mu)$ gives,   
\begin{equation} 
\begin{split}
\mu \frac{d}{d\mu} \frac{d\Gamma}{dz} =& \frac{M}{16\pi} \Bigl(z-\frac{\rc}{z}\Bigr)^2 |\mC_L(\mu)|^2\Biggl\{2~\mathrm{Re} [\gamma_{\mathcal{C}}] \int^1_z \frac{dx}{x}
\bar{J}_n (x,\mu) \bar{S}(z/x,\mu) 
\\
&\qquad + \int^1_z \frac{dx}{x} \Biggl[\Bigl(\mu \frac{d}{d\mu} \bar{J}_n (x,\mu) \Bigr) \bar{S}(z/x,\mu)+\bar{J}_n (x,\mu) \Bigl(\mu \frac{d}{d\mu} \bar{S}(z/x,\mu)\Bigr)\Biggr]\Biggr\},
\\
\label{sdidecnlo} 
=& \frac{M}{16\pi} \Bigl(z-\frac{\rc}{z}\Bigr)^2 |\mC_L(\mu)|^2
\Bigl[2~\mathrm{Re} [\gamma_{\mathcal{C}}^{\mathrm{LO}}] \cdot  \delta (1-z) + 
\Bigl(\gamma_J^{\mathrm{LO}} (z,\mu) + \gamma_S^{\mathrm{LO}} (z,\mu)\Bigr)\Bigr]+ \mO(\alpha_s^2),
\end{split}
\end{equation} 
where $\gamma_J$ and $\gamma_S$ are the anomalous dimensions for $\bar{J}_n$ and $\bar{S}$,
\begin{align}
\label{RGEJ}
\mu \frac{d}{d\mu} \bar{J}_n (x,\mu) &= \int^1_x \frac{dz}{z} \gamma_J(z,\mu) \bar{J}_n (x/z,\mu), 
\\
\label{RGES}
\mu\frac{d}{d\mu} \bar{S} (x,\mu) &= \int^1_x \frac{dz}{z} \gamma_S (z,\mu) \bar{S} (x/z,\mu).
\end{align}
At the lowest order in $\alpha_s$ they are given by 
\begin{align}
\label{gjLO} 
\gamma_{J}^{\mathrm{LO}} (z,\mu) &= \frac{\alpha_s}{\pi}C_F \Biggl[\delta(1-z) \Bigl(2\ln\frac{\mu^2}{Q^2} +\frac{3}{2}\Bigr) - \frac{2}{(1-z)_+}\Biggr], \\
\label{gsLO}
\gamma_{S}^{\mathrm{LO}} (z,\mu) &= \frac{\alpha_s}{\pi}C_F \Biggl[\delta(1-z) \Bigl(-\ln\frac{\mu^2}{M^2} +1 \Bigr) + \frac{2}{(1-z)_+}\Biggr].
\end{align} 
From~(\ref{anomalC}), (\ref{gjLO}), and (\ref{gsLO}) it then follows immediately that Eq.~(\ref{sdidecnlo}) vanishes at  $\mO(\alpha_s)$. 

\subsection{Resummed result for the differential decay rate near the end-point}

The factorized result in Eq.~(\ref{didec1}) still contains large logarithms. We resum these by RG evolving $|\mC_L|^2$, $\bar{J}_n$, and $\bar{S}$ from the factorization scale $\mu_F$ down to the respective ``typical scales'' for each of the three quantities. The RG evolution then automatically resums the large logarithms and exponentiates them. 
Here ``the typical scale'' denotes the scale at which the logarithms in the expressions for $|\mC_L|^2$, $\bar{J}_n$, and $\bar{S}$ are minimized. 
The typical hard scale for $|\mC_L|^2$ can be chosen as $\mu_H \sim M$. On the other hand,  Eqs.~(\ref{nlojet}) and (\ref{nlosoft}) imply that we can choose $\mu_J \sim M(1-\rc)^{1/2}(1-z)^{1/2}$ and $\mu_S \sim M(1-z)$ for the jet and soft functions, respectively. 

We perform the resummation to NNLL accuracy, counting large logarithms to be of $\mc{O}(1/\as)$.
For resummation to NNLL accuracy, we express the anomalous dimension of each factorized part as follows:
\begin{align}
\label{gc}
\gamma_{\mC} &= A_{\mC} \Gamma_C \ln \frac{\mu^2}{M^2(1-\rc)^2} +\hat{\gamma}_{\mC}, \\
\label{gj}
\gamma_{J} (z) &= \delta(1-z) \Bigl[A_{J} \Gamma_C \ln \frac{\mu^2}{M^2(1-\rc)} +\hat{\gamma}_{J}\Bigr] 
-\kappa_J  \frac{A_J\Gamma_C}{(1-z)_+} \ , \\
\label{gS}
\gamma_{S} (z) &= \delta(1-z) \Bigl[A_{S} \Gamma_C \ln \frac{\mu^2}{M^2} +\hat{\gamma}_{S}\Bigr] 
-\kappa_S  \frac{A_S\Gamma_C}{(1-z)_+} \ .
\end{align}
From Eqs.~(\ref{anomalC}), \eqref{gjLO} and \eqref{gsLO} we extract $\{A_{\mC},A_J,A_S,\kappa_J,\kappa_S\} = \{-1,2,-1,1,2\}$. 
$\Gamma_C$ is the cusp anomalous dimension~\cite{Korchemsky:1987wg,Korchemskaya:1992je}, which can be expanded as $\sum_{k=0} \Gamma_{k} (\alpha_s/4\pi)^{k+1}$. The first two coefficients in the expansion are,
\begin{equation} 
\label{cuspcoef} 
\Gamma_0 = 4 C_F,~~~\Gamma_1 = 4C_F \Bigl[\Bigr(\frac{67}{9} - \frac{\pi^2}{3} \Bigr)C_A - \frac{10}{9} n_f\Bigr],
\end{equation}  
where $C_A=N_c=3$ is the number of colors and $n_f$ is the number of flavors. The three loop coefficient $\Gamma_2$ reads~\cite{Moch:2004pa}
\begin{align}
\Gamma_2 =& 4C_F \Biggl[C_A^2\Bigl(\frac{245}{6}-\frac{134\pi^2}{27}+\frac{11\pi^4}{45}+\frac{22}{3}\zeta(3)\Bigr)+C_A n_f \Bigl(-\frac{219}{27}+\frac{20\pi^2}{27}-\frac{18}{3}\zeta(3)\Bigr) \nonumber \\
&~~+C_F n_f\Bigl(-\frac{55}{6}+8\zeta(3)\Bigr)-\frac{4}{27}n_f^2 \Biggr] \ .
\end{align}

The noncusp anomalous dimensions in Eqs.~\eqref{gj} and \eqref{gS} can be expanded as 
$\hat{\gamma}_{f=J,S}=\sum_{k=0} \hat{\gamma}_{f,k} (\alpha_s/(4\pi))^k$. From Eqs.~\eqref{gjLO} and \eqref{gsLO}, the leading coefficients are given as 
\be
\hat{\gamma}_{J,0} = 6C_F,~~~\hat{\gamma}_{S,0} = 4C_F.
\ee
The two loop coefficients required for NNLL accuracy are given by~\cite{Neubert:2004dd}
\begin{align}
\hat{\gamma}_{J,1} =& -2C_F\Biggl[C_F\Bigl(-\frac{3}{2}+2\pi^2-24 \zeta(3)\Bigr)+C_A \Bigl(-\frac{3155}{54}+\frac{22\pi^2}{9}+40\zeta(3)\Bigr) \nonumber \\
&~~~+n_f\Bigl(\frac{247}{27}-\frac{4\pi^2}{9}\Bigr)+2\beta_0 (7-2\pi^2)\Biggr] \ ,\\
\hat{\gamma}_{S,1} =& 16C_F\Biggl[C_A\Bigl(-\frac{37}{108}+\frac{\pi^2}{144} +\frac{9}{4} \zeta(3)-\frac{1}{6}\Bigr)-n_f \Bigl(\frac{1}{54}+\frac{\pi^2}{72}\Bigr) \Biggr] \ ,
\end{align}
where $\beta_0$ is the first coefficient of QCD beta function. The $\hat{\gamma}_{\mC}$ in Eq.~\eqref{gc} can be written as $\hat{\gamma}_{\mC} = -\hat{\gamma}_{J} -\hat{\gamma}_{S}$ from the fact that the the differential decay width is scale independent. 

Performing the RG evolutions using Laplace 
transform~\cite{Neubert:2005nt,Becher:2006nr} leads to the resummed result near the end-point at NNLL accuracy as 
\begin{equation}
\begin{split}
\label{Resumr}
\frac{d\Gamma_{\rm res}(\tilde{q} \to \chi^0 X_{q_L} )}{dz} =&
\frac{M}{16\pi}  (1-\rc)^2 \exp[\mathcal{M} 
(\mu_{H},\mu_{J},\mu_{S})]|B_L(\mu_{H})|^2 |C_L(M,\rc,\mu_{H})|^2  
 \\
&\times \tilde{J} \Bigl[\ln\frac{\mu_{J}^2}{M^2(1-\rc)}-\partial_{\eta}\Bigr] 
\tilde{S} \Bigl[\ln\frac{\mu_S^2}{M^2}-2\partial_{\eta}\Bigr]
\frac{e^{-\gamma_E\eta}}{\Gamma(\eta)} (1-z)^{-1+\eta},
\end{split}
\end{equation}
For integrated quantities such as the total decay width the NNLL accuracy requires  jet and soft functions, $\tilde {J}$ and $\tilde {S}$, to be calculated at ${\mathcal O}(\as)$ (see, e.g., Table 1 in Ref.~\cite{Becher:2006mr}). For decay width distributions, such as $d\Gamma/dz$, the log enhanced ${\mathcal O}(\as^2)$ terms in $\tilde {J}_n$ and $\tilde {S}$ need to be included as well~\cite{Almeida:2014uva} (see, e.g., Table 6 in Ref.~\cite{Almeida:2014uva}). To the required order the two functions are
\begin{align}
\tilde{J} [L] =& 1+\frac{\alpha_s C_F}{2\pi} \Bigl(\frac{7}{2}-\frac{\pi^2}{3}+\frac{3}{2}L+L^2\Bigl)+\cdots, \label{JJ} \\
\tilde{S} [L] =& 1+\frac{\alpha_s C_F}{2\pi} \Bigl(-\frac{5\pi^2}{12}+L-\frac{1}{2} L^2\Bigr)+\cdots.  \label{SS} 
\end{align}
where the ellipses denote the $\mc{O}(\as^2)$ terms, which are given in Appendix \ref{two_loop}. For the NLL result, we only keep the first two terms in Eq.(\ref{JJ}) and (\ref{SS}). 
%

The exponantiation factor in Eq.~\eqref{Resumr} is given by 
\begin{equation}
\begin{split}
\mathcal{M} (\mu_{H},\mu_{J},\mu_{S}) = &~ 2 S_{\Gamma}(\mu_H,\mu_J) -2 S_{\Gamma}(\mu_J,\mu_S)
+\ln\frac{\mu_H^2}{M^2(1-\rc)^2} a[\Gamma_C](\mu_H,\mu_J) 
 \\
&~ -\ln\frac{\mu_H^2}{M^2} a[\Gamma_C](\mu_J,\mu_S) + a[\hat{\gamma}_J](\mu_H,\mu_J)
+a[\hat{\gamma}_S](\mu_H,\mu_S),
\end{split}
\end{equation}
with the Sudakov factor $S_{\Gamma}$ and the evolution function $a[f]$ defined as 
\begin{equation} 
S_{\Gamma}(\mu_1,\mu_0) = \int^{\alpha_1}_{\alpha_0} \frac{d\alpha}{b(\alpha)} \Gamma_C(\alpha) \int^{\alpha}_{\alpha_1} \frac{d\alpha'}{b(\alpha')},\qquad
a[f] (\mu_1,\mu_0) = \int^{\alpha_1}_{\alpha_0} \frac{d\alpha}{b(\alpha)} f(\alpha).
\end{equation}
Here $\alpha$ and $\alpha_{0,1}$ denote $\alpha_s(\mu)$ and $\alpha_s(\mu_{0,1})$, while $b(\alpha_s)=d\alpha_s/d\ln\mu$ is the QCD beta function. 
To NNLL accuracy $S_{\Gamma}$ and $a[\Gamma_C]$ are given by~\cite{Becher:2006mr}
\begin{align} 
\begin{split}
S(\mu_1,\mu_0) =& \frac{\Gamma_0}{4\beta_0^2} \Biggl\{\frac{4\pi}{\alpha_s(\mu_1)}\Bigl(1-\frac{1}{r} - \ln r\Bigr)+ \Bigl(\frac{\Gamma_1}{\Gamma_0}-\frac{\beta_1}{\beta_0}\Bigr)(1-r+\ln r)+ \frac{\beta_1}{2\beta_0}\ln^2 r  \\
\label{SNNLL}
&+\frac{\alpha_s(\mu_1)}{4\pi}\Biggl[\Bigl(\frac{\beta_1\Gamma_1}{\beta_0\Gamma_0}-\frac{\beta_2}{\beta_0}\Bigr)(1-r+r\ln r)+\Bigl(\frac{\beta_1^2}{\beta_0^2}-\frac{\beta_2}{\beta_0}\Bigr)(1-r)\ln r  \\
& - \Bigl(\frac{\beta_1^2}{\beta_0^2}-\frac{\beta_2}{\beta_0}-\frac{\beta_1\Gamma_1}{\beta_0\Gamma_0}-\frac{\Gamma_2}{\Gamma_0}\Bigr)\frac{(1-r)^2}{2}\Biggr]\Biggr\}, 
\end{split}
 \\
a[\Gamma_C] (\mu_1,\mu_0) =& \frac{\Gamma_0}{2\beta_0}\Bigl[\ln r +\Bigl(\frac{\Gamma_1}{\Gamma_0}-\frac{\beta_1}{\beta_0}\Bigr)\frac{\alpha_s(\mu_0)-\alpha_s(\mu_1)}{4\pi}\Bigr], \label{aGamma}
\end{align}
wehere $r=\alpha_s(\mu_0)/\alpha_s(\mu_1)$.
Finally, the evolution parameter $\eta$ in Eq.~\eqref{Resumr} is defined as $\eta =2a[\Gamma_C](\mu_J,\mu_S)$. It is  positive since $\mu_J>\mu_S$.
For the NLL result, we only keep the first line of Eq. (\ref{SNNLL}), and the first term in Eq. (\ref{aGamma}).

\section{Decay distribution in the full range}
\label{FullRange}
Even though the decay distribution $d\Gamma/dz$ in the region $z\to 1$ is the dominant contribution to the total decay width, it is useful for phenomenological analyses to obtain the decay distribution in the full range of $z$, while retaining the $\mO(1-z)$ corrections. The expression for $d\Gamma/dz$ away from $z\to 1$ should be obtained using full QCD. Away from the end-point region the gluon emissions  are hard so that the total invariant mass of final state jets can be comparable to $M$. 

To perform the calculation of $\mT_{L}$ in Eq.~(\ref{TE}) in full QCD we introduce the structure function $W_L (z,\rc,M)$, 
\beq
\mT_L (z,\rc, M) = 2\pi|B_L|^2 W_L (z,\rc,M).
\eeq
 The $W_L$ is thus given by 
\begin{equation} 
\label{stu}
W_L (z,\rc,M) = \frac{1}{2\pi}\int d^4 z e^{-ip_{\chi}\cdot z} 
\langle \tilde{q} |  \tilde{q}^{*\alpha} (P_L q^{\alpha})_a (z) (\bar{q}^{\beta} P_R)_b \tilde{q}_L^{\beta}(0) |\tilde{q}\rangle (\fms{p}_{\chi} - m_{\chi})_{ba},
\end{equation}
where $p_{\chi}$ is the momentum of the neutralino (the expression for it in terms of $z$ and $\rc$ is given in Eq.~(\ref{pchi})). The differential decay rate is then 
\begin{equation} 
\label{fulldecd}
\frac{d\Gamma}{dz}
(\tilde{q} \to \chi^0 X_{q_L}) = \frac{M}{16\pi} |B_L (\mu) |^2 \frac{1}{z} \Bigl(z-\frac{\rc}{z}\Bigr)^2 W_L (z,\rc,M,\mu). 
\end{equation}
At tree level we have simply $W_L=\delta(1-z)$.

\begin{figure}[t]
\begin{center}
\includegraphics[height=7.5cm]{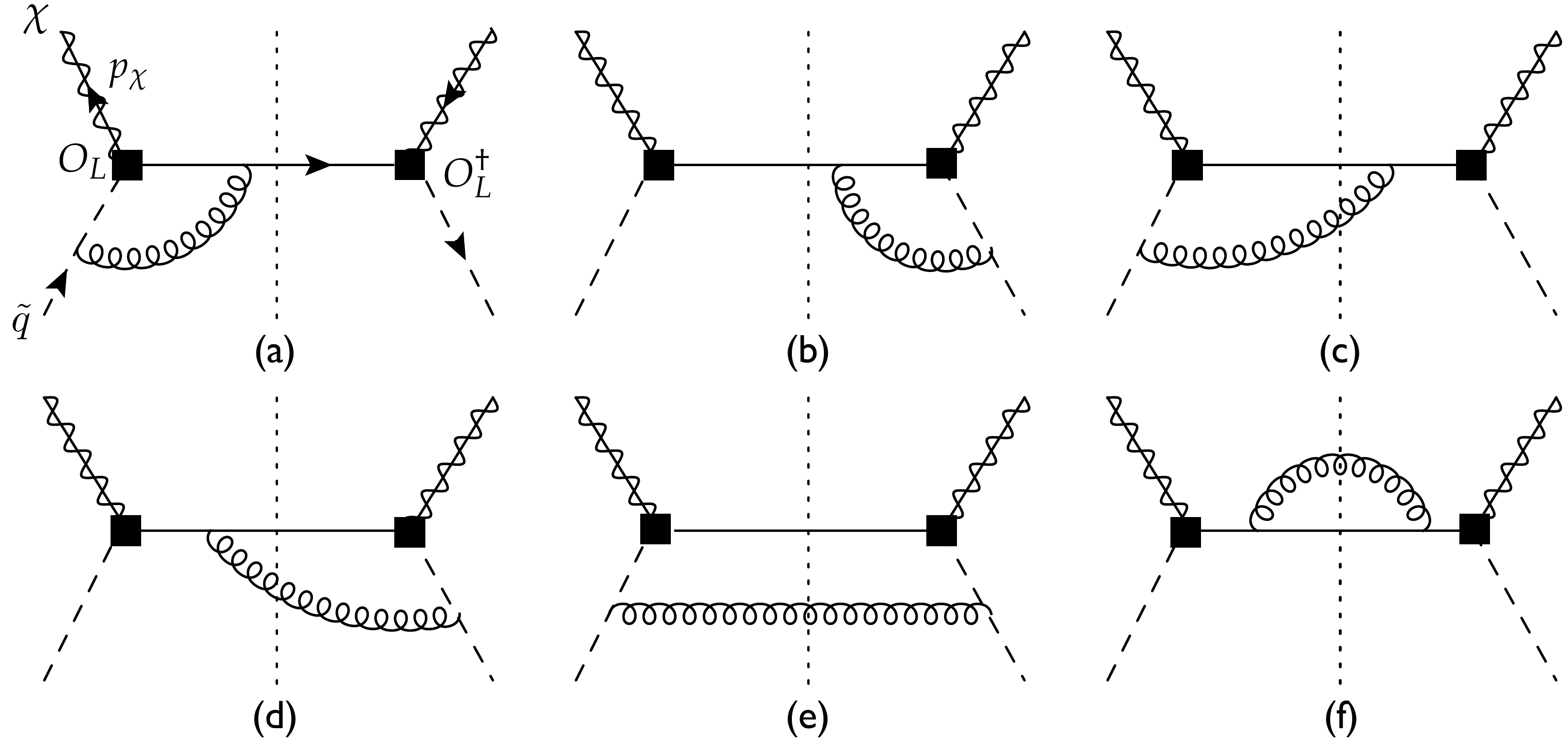}
\end{center}
\vspace{-0.3cm}
\caption{\label{fig2} \baselineskip 3.0ex 
The NLO Feynman diagrams  for $\tilde{q} \to \chi X_{q_L}$. The dashed lines at the center of each diagram denotes the discontinuity cut for forward scattering amplitudes. 
} 
\end{figure}

To obtain the NLO expression for $W_L$ we computed the Feynman diagrams shown in Fig.~\ref{fig2} as well as self energy diagrams for the squark and the light quark. As a result, one loop corrections to $W_L$ in $\overline{\rm MS}$ scheme are given as   
\begin{equation} 
\begin{split}
\label{Wonel}
W^{(1)}(z,\rc,M,\mu) =& \frac{\alpha_s C_F}{2\pi}\Biggl\{ \delta(1-z) \Biggl[ \frac{3}{2}\ln \frac{\mu^2}{M^2(1-\rc)^2} +\frac{1}{4} \ln^2 (1-\rc) + 2 \mr{Li}_2 \Bigl(\frac{\rc}{\rc-1}\Bigr) \\
&-2\rc H_1(\rc)-\frac{7}{2}\rc H_2 (\rc)+ \frac{5}{4}-\frac{\pi^2}{3}\Biggr]+\frac{\rc}{z-\rc} \\
&+\frac{1}{(1-z)_{\Delta}} \frac{z}{z-\rc} \Bigl[-4(1-\rc)+\frac{1}{2}\Bigl(z - \frac{\rc}{z}\Bigr)\Bigr]  \\
&-2\frac{z(1-\rc)(2-z-\rc/z)}{z-\rc} \Bigl[g(z,\rc)\Bigr]_{\Delta} \Biggr\} \ . 
\end{split} 
\end{equation}
Here we introduced the so called `delta distribution', $[\ldots]_\Delta$, in order to deal with the infrared (IR) singularity as $z\to 1$. The definition and some useful properties of the $\Delta$ distribution are given in Appendix~\ref{AB}.  For a region of integration that is, as in our case, over $z \in [\sqrt{r_{\chi}},1]$ (rather than over the interval $[0,1]$) the introduction of a $\Delta$ distribution shortens the expressions compared to the standard plus distribution. 
 In Eq.~\eqref{Wonel} the functions $H_{1,2}(\rc)$ and $g(z,\rc)$ are given by, 
\begin{align} 
\label{H1e}
H_1(\rc) &= \int^1_{\sqrt{\rc}} \frac{dz}{z} \frac{(2-\rc-\rc/z)(1-2\rc/z+\rc/z^2)}{(z-\rc/z)(1-\rc/z)^2} \ln \frac{1-z}{1-\rc/z},\\
\label{H2e}
H_2(\rc) &= \int^1_{\sqrt{\rc}} \frac{dz}{z}\frac{1-2\rc/z+\rc/z^2}{(1-\rc/z)^2}
=\frac{1}{\sqrt{\rc}}+ 2\coth^{-1} (1+2\sqrt{\rc}) - \frac{\ln(1+\sqrt{\rc})}{x},\\
\label{gze}
g(\rc,z) &= \frac{z}{(1-z)(z^2-\rc)} \ln \frac{1-z}{1-\rc/z}.
\end{align}

The anomalous dimension $\gamma_W (z)$, controlling the RG evolution of structure functions, 
\begin{equation} 
\label{RGEW} 
\mu \frac{d}{d\mu} W_L (M,\rc, z,\mu) = \int^1_z \frac{dx}{x} \gamma_W(x) W_L (M, \rc,z/x,\mu) 
\end{equation} 
is given by
\beq
\label{eq:GammaWz}
\gamma_W (z) = 3\frac{\alpha_s C_F}{2\pi} \delta(1-z) +\mO(\alpha_s^2).
\eeq
Using the fact that $d\Gamma/dz$ is scale-invariant then gives the anomalous dimension for $B_L(\mu)$ as $\gamma_B = -3\alpha_s C_F/(2\pi)+\mO(\alpha_s^2)$, which, as expected, is the same result as given in Eq.~(\ref{gammaB}). 

Finally, we combine our results for the differential decay rates in the full $z$ range and near the end-point, $z\to1$, to obtain the decay distribution at NNLL+NLO,\footnote{One may wish to consider other matching schemes that  turn off smoothly the resummation effects in the region far away from the end-point. In our case the role of the smooth matching is performed by the running of jet and soft scales which are respectively given as 
$\mu_J^0 = M\sqrt{(1-r_{\chi})(1-z)}$ and $\mu_S^0 = M(1-z)$, where $r_{\chi}=m_{\chi}^2/M^2$. 
 While the prescription in Eq. \eqref{RG_improved} does not recover exactly the NLO result anywhere in the physical region $z\geq \sqrt{r_{\chi}}$, this treatment does suffice for our purposes. First of all, alternative prescriptions that have  $\mu_H=\mu_J=\mu_S$ at $z=0$ similarly do not lead to perfectly smooth matchings anywhere in the physical region for  $z$. More importantly, 
the two decay distributions with and without resummations, ${d\Gamma_{\rm res}}/{dz}$ and ${d\Gamma_{\mathrm{NLO}}^f}/{dz}$,  are completely dominated by the end-point region and have only negligible contributions from the rest of the $z$ range. We therefore expect only numerically subleading corrections from alternative matching prescriptions relative to the results using Eq.~\eqref{RG_improved}.
 }
\begin{equation} 
\label{RG_improved} 
\frac{d\Gamma_{\rm full}}{dz} (\tilde{q}_L \to \chi^0 X_q) = \frac{d\Gamma_{\rm res}}{dz} + \frac{d\Gamma_{\mathrm{NLO}}^f}{dz} - \frac{d\Gamma^{f}_{E}}{dz}.
\end{equation}
The first and the second terms on the right-hand side are the resummed result near the end-point, Eq.~(\ref{Resumr}), and the NLO result for the full $z$ range, Eq.~(\ref{fulldecd}), respectively. The double counting of contributions between the two terms is removed by the third term on the right-hand side of Eq.~\eqref{RG_improved}, i.e., the $d\Gamma^f_E/dz$. The expression for $d\Gamma^f_E/dz$ follows from $d\Gamma_{\rm res}/dz$ by identifying the multiple scales as $\mu_{H} = \mu_J =\mu_S = M$.

\section{Phenomenological study}
\label{sec:pheno}

In this section, we present a detailed study of NLO, NLL+NLO, and NNLL+NLO predictions for the $\tilde q\to q \chi$ decay. In section~\ref{sec:results} we show our results for the normalized differential width distributions as well as the total decay widths, and discuss the impact of soft gluon resummations on the NLL+NLO and NNLL+NLO results. In section~\ref{sec:simulation}, we perform a numerical analysis at NLL+NLO and NNLL+NLO accuracies for the decays of pair-produced squarks in a linear $e^{+}e^{-}$ collider based on $\sqrt{s} = 3$ TeV Compact Linear Collider (CLIC). Since the decay topology of a squark can be significantly altered by higher-order corrections, it is necessary to scrutinize these effects for the precise measurements of a squark and neutralino masses, which is an important part of the CLIC physics program.

\subsection{Differential width distributions and total widths}
\label{sec:results}

The resummed results that we calculated at NNLL+NLO and NLL+NLO accuracies in Eq.~(\ref{Resumr}) depend on the choices of scales, $\mu_H$, $\mu_J$, and $\mu_S$. To illustrate the scale dependences, we independently vary the $\mu_{i}$, $i=H,J,S$, between $2\mu^0_{i}$ and $\mu^0_{i}/2$, 
where $\mu_{i}^0$ are the default choices of the hard, jet, and soft scales. 
We take $\mu_H^0 = M$ for the default hard scale. 
The default jet and soft scales are chosen as the running scales $\mu_J^0 = M\sqrt{(1-\rc)(1-z)}$ and $\mu_S^0 = M(1-z)$,  where $r_{\chi}=m_{\chi}^2/M^2$.
For $z\to1$ we would have $\mu_i^0\to0$ for these choices of running scales and therefore the IR Landau poles in the running of the strong coupling constants. 
In order to avoid this problem we adopt the following profile function for the soft scale,
\be
\label{spf} 
\mu_S^0=\mu_{S}^{pf} = 
\left\{
\begin{array}{rl} 
& M(1-z)~~~\mr{if}~z \le z_0,  \\ 
& \mu_{\mr{min}} + a  M(1-z)^2 ~~~\mr{if}~z > z_0.  \end{array} \right.
\ee
Therefore we make the soft scale frozen as $\mu_{\mr{min}}$ as $z\to 1$. The parameters $a$ and $z_0$, where $1-z_0 \ll 1$,
are determined by $z_0=1 - 2 \mu_{\mr{min}} / M $ and $ a=M/ (4 \mu_{\mr{min}} )$ to ensure that $\mu_{S}^{pf}$ is smoothly continuous at $z_0$. 
We use $\mu_{\mr{min}} = 0.5$~GeV so that  $M(1-z_0)=1$~GeV.  
The impact of nonperturbative physics grows as $z$ becomes larger than $z_0$, and therefore $\mu_S^0$ goes from $1$ GeV to $0.5$ GeV. The precise choice of $\mu_{\mr{min}}$ and the estimation of its uncertainty would be possible from a nonperturbative model or from a fit to experimental data, if these become available. This is beyond the scope of this paper, where we focus on perturbative resummation effects near the end-point. 

We also modify the jet scale using the following profile function
\footnote{Eq.~(\ref{jpf}) relates only the default jet and soft scales. When we estimate the uncertainty due to the choice of scales, the jet and soft scales are varied independently around these central values.
} 
\beq
\label{jpf}
\mu_J^0=\mu_{J}^{pf} = \sqrt{M(1-\rc)\mu_{S}^{pf}}.
\eeq  

\begin{table}[t]
\begin{center}
\scalebox{0.88}{
\begin{tabular}{|c|c|c|cc|cc|}
\hline
~$M_{\chi}$ (GeV) ~ &  ~ LO (GeV) ~              &  ~NLO(GeV)~                        & ~NLL+NLO (GeV)  ~      &  ~ unc.$(\%)$ ~  & ~NNLL+NLO (GeV)  ~         &  ~ unc.$(\%)$ ~    \\ \hline \hline
                     1000     &                       $7.93$   &     $7.87$                                &  $7.45$                    &   $\pm 18.13$             &  $8.58$                           &   $\pm 14.81$                  \\
                       900     &                        $10.90$ &    $10.85$                              &         $9.92$                   &   $\pm 17.79$               &  $11.47$                           &   $\pm 14.20$                 \\
                       800 	&                        $13.96$ &    $13.90$                              &         $12.40$                  &   $\pm 17.76$               &  $14.38$                           &   $\pm 13.94$                \\
                       700  &                        $16.97$  &     $16.91$                               &        $14.79$                  &  $\pm 17.90$               &  $17.19$                           &   $\pm 13.90$                 \\
                       600  &                       $19.81$ &       $19.74$                               &        $17.01$                  &   $\pm 18.13$               &  $19.81$                           &   $\pm 13.98$                \\
                       500  &                       $22.39$ &       $22.30$                               &       $18.98$                   &    $\pm 18.39$              &  $22.15$                           &   $\pm 14.12$                 \\
                       400  &                        $24.62$  &      $24.50$                              &        $20.65$                  &   $\pm 18.66$               &  $24.14$                           &   $\pm 14.29$                 \\
                       300  &                        $26.43$ &       $26.29$                              &       $22.00$                   &   $\pm 18.91$               &  $25.73$                           &   $\pm 14.46$                 \\
                       200  &                        $27.76$ &       $27.62$                              &        $23.23$                  &   $\pm 18.93$               &  $27.16$                           &   $\pm 14.49$                 \\ \hline
\end{tabular}}
\end{center}
\caption{\baselineskip 3.0ex Total decay widths of a $1.45 \TeV$ squark based on the LO, fixed-order (NLO), NLL+NLO, and NNLL+NLO calculations, with the Wilson coefficient $B_L$ at the scale $M$ taken to be $B_L(M)=1$. Benchmark neutralino masses are varied in the interval $200<M_{\chi}<1000 \GeV$. The impact of scale uncertainties for NLL+NLO and NNLL+NLO predictions are shown as well.
}
\label{tab:WidthTable}
\end{table}

\begin{figure}[t]
\includegraphics[scale=0.83]{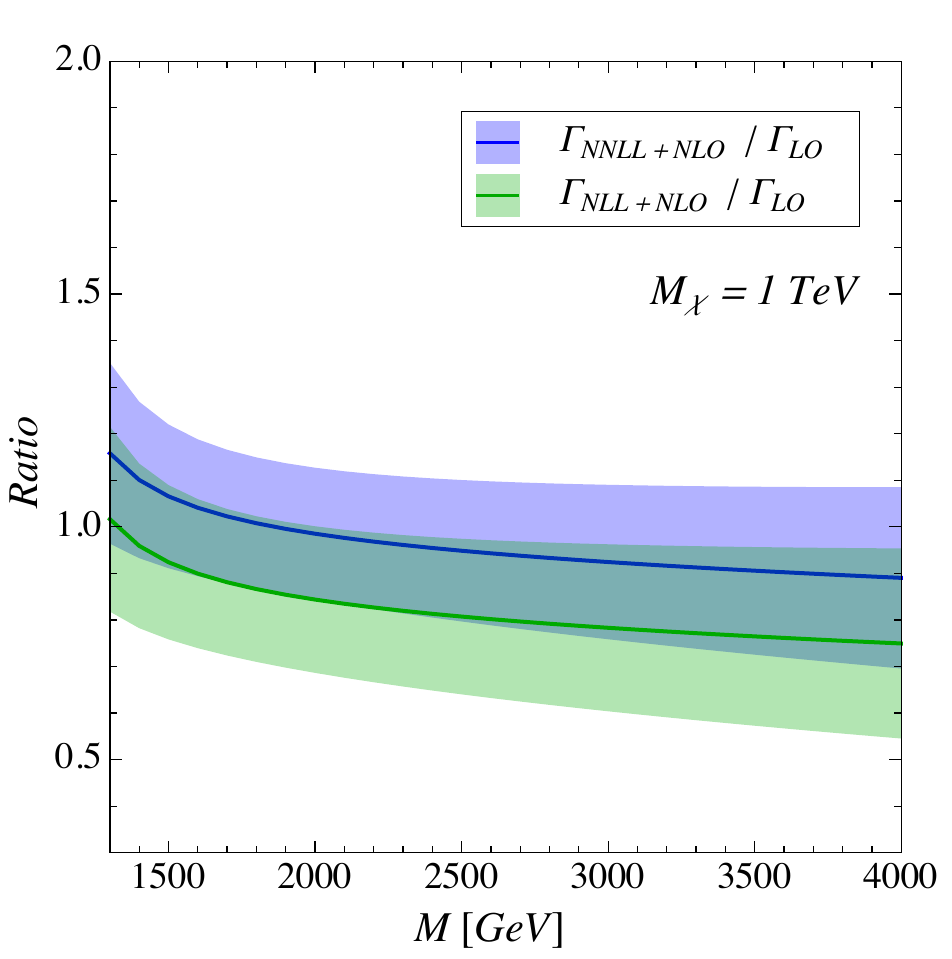}
\includegraphics[scale=0.83]{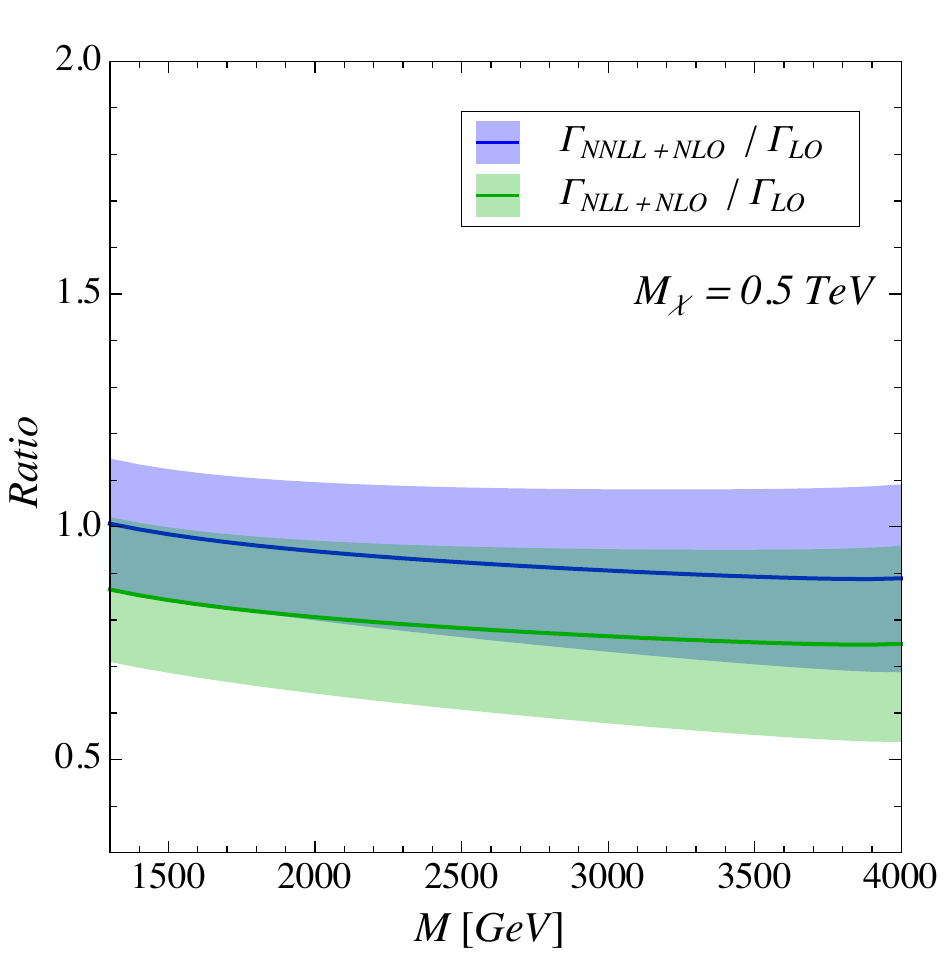}
\vspace{-0.3cm}
\caption{\label{fig:TotWidth}The ratios of NLL+NLO (green line) and NNLL+NLO (blue) total decay widths normalized to the LO result as a function of mass $M$. The neutralino masses are fixed to $M_{\chi}=1~(0.5)\TeV$ in the left (right) panel.
The details on hard, jet, and soft scale variations, giving the corresponding bands, are explained in Eq. (\ref{spf}) and Eq. (\ref{jpf}), with $\mu_{\mr{min}} = 0.5~\mr{GeV}$.
}
\end{figure}

\begin{figure}
\includegraphics[scale=0.7]{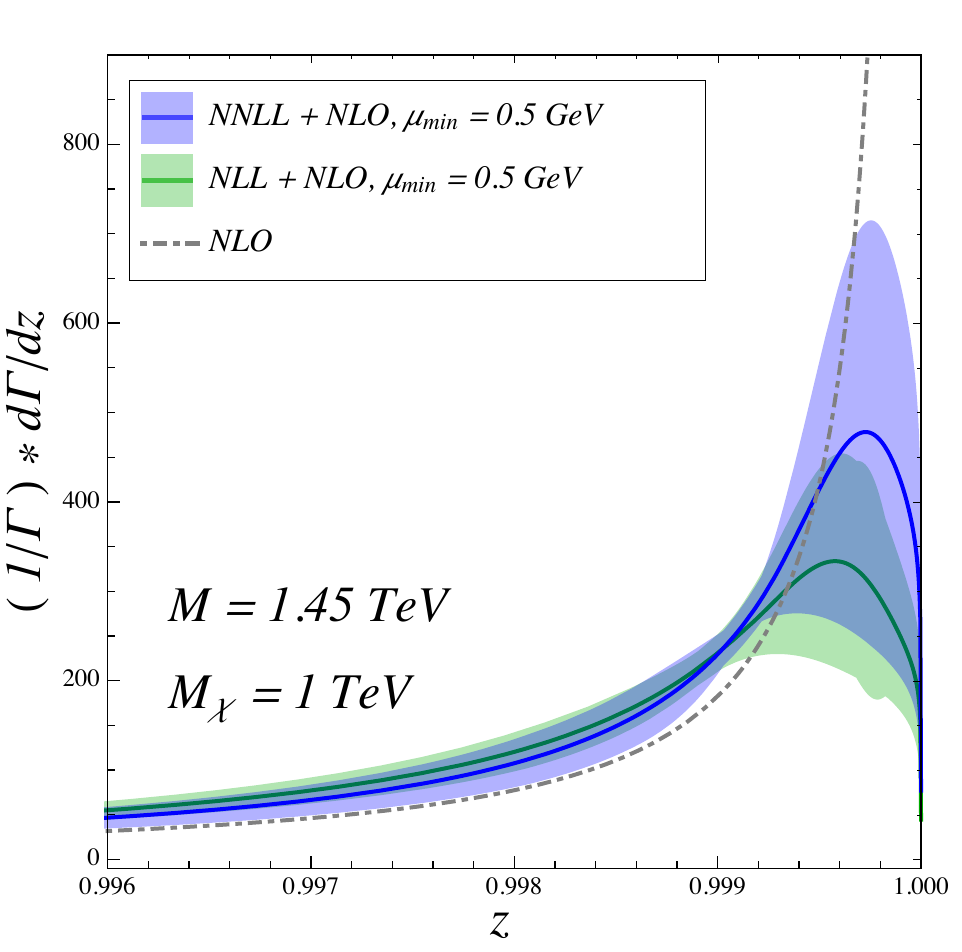}
\includegraphics[scale=0.7]{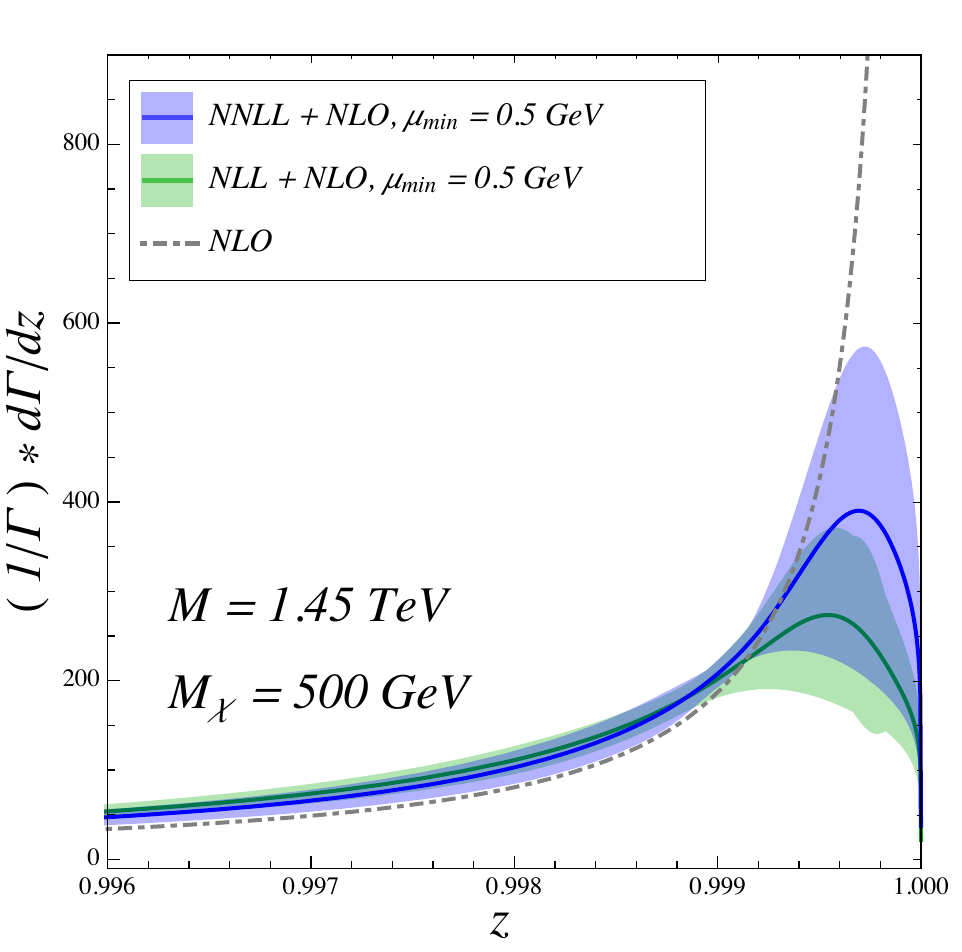} \\
\vspace{-0.3cm}
\caption{\label{fig:DW} \baselineskip 3.0ex Normalized NLL+NLO (green), NNLL+NLO (blue), and NLO (gray dash-dotted) differential decay width distributions, using $\mu_{\mr{min}} = 0.5~\mr{GeV}$. 
We fixed neutralino masses to (left panel) $M_{\chi}=1\TeV$ and (right panel) $M_{\chi} = 0.5\TeV$.}
\end{figure}

Table \ref{tab:WidthTable} shows the total decay widths of a squark with mass $M = 1.45 \TeV$ obtained at LO, NLO, NLL+NLO, and NNLL+NLO accuracies, with the Wilson coefficient $B_L$ normalized by $B_L(M)=1$ at the scale $M$. Benchmark neutralino masses are chosen in the interval $200<M_{\chi}<1000 \GeV$. The scale uncertainty of NLL+NLO prediction turns out to be $\sim 18\%$, which is improved to $\sim 14\%$ at NNLL+NLO. These are obtained by varying the hard, jet and soft scales in the range from $\mu_{H,J,S}^0/2$ to $2\mu_{H,J,S}^0$ each independently. Figure \ref{fig:TotWidth} shows the ratios of total decay widths of NLL+NLO and NNLL+NLO predictions with respect to the LO  as a function of mass $M$, while fixing (left) $M_{\chi} = 1 \TeV$ and (right) $M_{\chi} = 0.5 \TeV$. For both cases, scale uncertainties become larger as $M$ increases.

Normalized NLL+NLO, NNLL+NLO, and NLO differential decay width distributions using $\mu_{\mr{min}} = 0.5~\mr{GeV}$ are shown in Figure \ref{fig:DW}. The NLO distribution diverges in the region $z \rightarrow 1$, while the resummed NLL+NLO and NNLL+NLO distributions are regulated at the end-point. The central value of NLL+NLO distribution tends to be more broadened compared to the NNLL+NLO, but their overall size of uncertainties near the end-point region are similar with each other. 

Concerning the scale variations of the NLL+NLO and NNLL+NLO distributions, we find that the dominant uncertainties come from when we vary the soft scale down to $\mu^0_S/2$, while other scale variations give quite small uncertainties. Note that $\mu^0_S$ varies from $5.8$ GeV to $1.45$ GeV when $z$ is changed from $z=0.996$ to $z=0.999$, and reaches $\mu_{\rm min}=0.5$ GeV at the very right end of panels in Fig. \ref{fig:DW}. The large variation still present at NNLL for $z>0.999$ can thus be traced to this very low soft scale that is reached at the very end-point of the spectrum.

\subsection{Precision studies of squarks and neutralinos at CLIC}
\label{sec:simulation}
The squark decay, $\tilde q\to q \chi$, results in a two-body final state at LO, which at  higher orders becomes a multi-body final state due to additional hard or soft QCD radiation. 
This can potentially affect the methods for precise measurements of squark and neutralino masses. As an illustration we take the impact of QCD corrections on such measurements at  CLIC~\cite{Assmann:2000hg, Battaglia:2004mw, Aicheler:2012bya, Abramowicz:2016zbo}, a future linear $e^{+}e^{-}$ collider designed to provide collision energies up to 3 TeV. If supersymmetric particles are light enough to be produced at such machine, CLIC will provide a platform for precision studies where their properties could be determined with considerable accuracy~\cite{Dannheim:2012rn, Battaglia:2013bha, Munnich:2012, Barklow:2011, Simon:2015pza}. In the phenomenological analysis we focus exclusively on the impact of QCD radiations in the squark decay. For a realistic study other important effects, in particular the initial state QED radiation, that results in the reduced effective $e^+e^-$ collision energy, need to be included. 

For pair-produced squarks that decay into light quarks and neutralinos,
\bea
\label{eq:process} 
	e^{+} ~ e^{-} \rightarrow \tilde{q} ~ \tilde{q}^{*} \rightarrow q ~ \chi ~ \overline{q} ~\chi   \, ,
\eea
an interesting technique to simultaneously measure squark and neutralino masses, is to search for the edges in the event distributions. We will discuss two such methods, i) based on edges in energy distribution of the light quark jets, $E_1+E_2$,  and ii) a method based on the kinematic variable $M_C$ \cite{Tovey:2008ui}. 
The numerical analysis is based on LO version of \amc ~\cite{Alwall:2011uj, Alwall:2014hca} with \verb|PYTHIA 6| \cite{Sjostrand:2006za} showering, but no hadronization, which was used to generate the event chain in \eqref{eq:process}, utilizing the Minimal Supersymmetric Standard Model (MSSM) implementation from~\cite{Christensen:2009jx, Duhr:2011se}. 
The {\tt PYTHIA} events were clustered with the \verb|FastJet| \cite{Cacciari:2011ma} implementation of the anti-$k_T$ algorithm \cite{Cacciari:2008gp}, taking $r = 0.4$ for the cone size. For events to pass the selection cuts we require at least two jets with $p_T > 50 \GeV$. To obtain the NLL+NLO and NNLO+NLO (including two-loop-log terms) samples we reweight {\tt PYTHIA} events, on an event-by-event basis, according to the $d\log\Gamma/dz$ normalized differential distributions in Figure \ref{fig:DW} for each of the decay chains. We first rewrite the variable $z$ in Lorentz-invariant form
\begin{equation}
z = \frac{x + \sqrt{x^2 - 4 r_{ \chi}}}{2}   \;\;\;\;\; \text{with}\;\;\; x = 2 \; \frac{p_{\chi} \cdot p_{\tilde q}}{M^2}  \;,
\label{Eq:z}
\end{equation}
where $p_{\tilde q}$ and $p_{\chi}$ are four-momenta of a squark and a neutralino respectively\footnote{We keep track of the event history to access this information.}. We plug two $z$ values (from two decay chains) into the normalized NLL+NLO and NNLO+NLO distributions in Figure \ref{fig:DW} using $\mu_{\mr{min}} = 0.5~\mr{GeV}$ to obtain probabilities. Then we multiply the two probabilities to obtain the weight for each event. In this way, the simulated events acquire the correct NLL+NLO and NNLO+NLO distributions in $z$ variable, but are only approximately NLL+NLO and NNLO+NLO in the other phase space variables. The NLO samples, on the other hand, cannot be obtained using the same reweighting method. Since the NLO distributions in Figure \ref{fig:DW} diverge at $z = 1$, the probabilistic interpretation of the differential width distributions is not well-defined. As a result we do not include reweighted NLO distributions. We derive results for two benchmarks, setting squark mass to $M = 1.45 \TeV$, while taking the lightest neutralino mass to be $M_{\chi} = 1\TeV$ or $0.5 \TeV$, and assume a negligible squark decay width. In this study, the beamstrahlung, initial state radiation, and detector effects are not included.

\begin{figure}[!]
\begin{center}
\begin{tabular}{cc}
\hspace{-10pt} \includegraphics[scale=0.4]{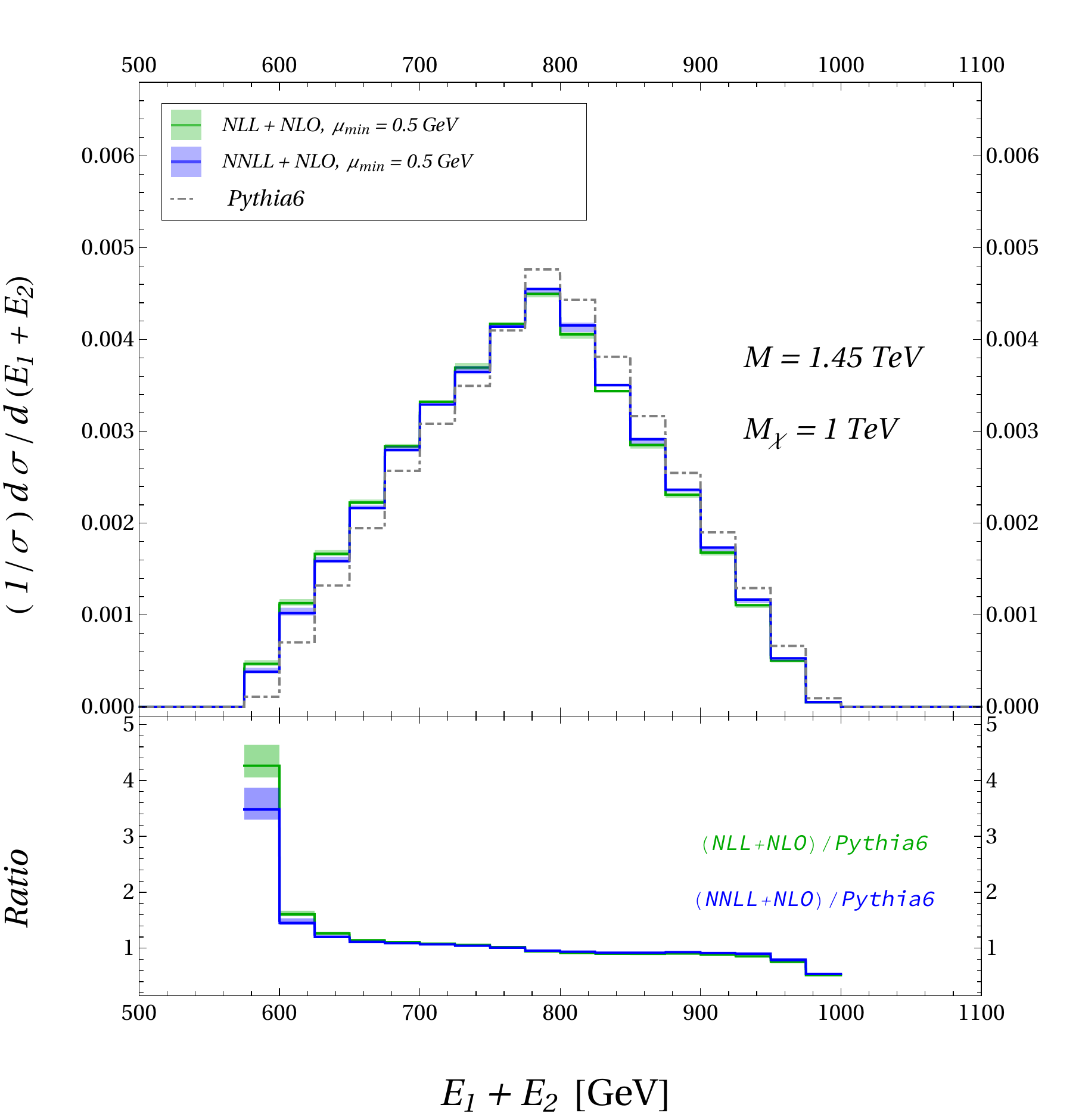} &
\hspace{+10pt} \includegraphics[scale=0.4]{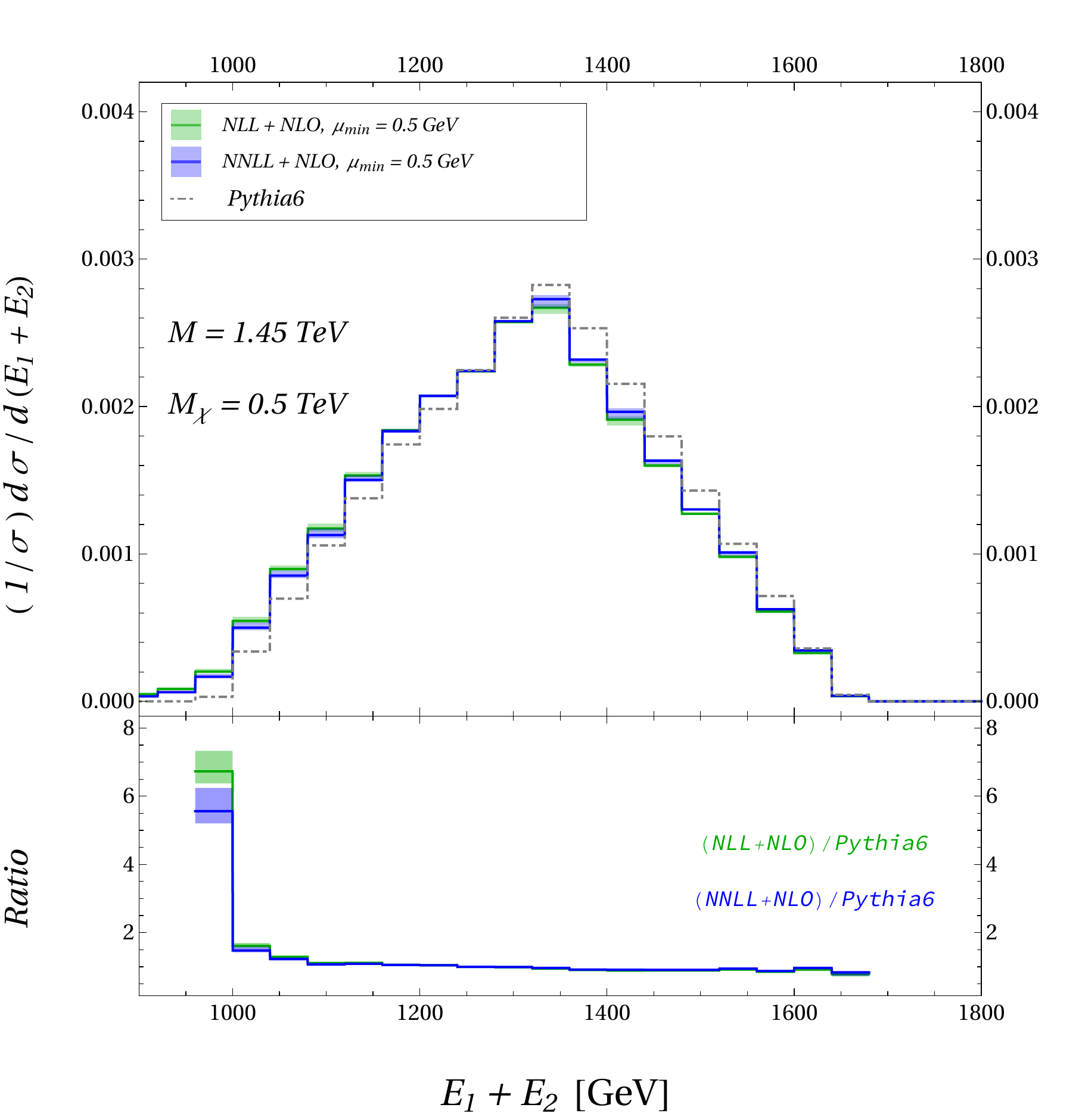} 
\vspace{-0.5cm}
\end{tabular}
\end{center}
\caption{\label{fig:EPlots} \baselineskip 3.0ex 
Distributions for the sum of energies of the first two hardest jets for $M_{\chi}= 1 \TeV$ (left)  and  $M_{\chi}= 500 \GeV$ with fixed  $M = 1.45 \TeV$ (right).
} 
\end{figure}

For two body squark decays, \eqref{eq:process}, the minimal and maximal light quark energy are directly related to $M$ and $M_{\chi}$ \cite{Feng:1993sd, Frank_Simon}
\begin{align} 
E_{q, {\rm max}}  =& \frac{\sqrt{s}}{4} \Bigg(  1 - \frac{M^2_{\chi}}{M^2}   \Bigg) \Bigg( 1 + \sqrt{ 1 - \frac{4 M^2}{s} }  \Bigg)  \; , \label{Eq:E1} \\
E_{q, {\rm min}}  =& \frac{\sqrt{s}}{4} \Bigg(  1 - \frac{M^2_{\chi}}{M^2}   \Bigg) \Bigg( 1 - \sqrt{ 1 - \frac{4 M^2}{s} }  \Bigg)   \;, \label{Eq:E2}
\end{align} 
and thus in our case $E_1+E_2\in[ 2 E_{q,{\rm min}}, 2 E_{q,{\rm max}}]$, neglecting the small squark boosts in the lab frame. At LO the $E_1+E_2$ distributions start at $0.59\TeV$ and $0.98\TeV$, for $M_\chi=0.5\TeV$ and $M_\chi=1\TeV$, respectively. In Figure \ref{fig:EPlots}, on the other hand, the NLL+NLO and NNLL+NLO $E_1 + E_2$ distributions extend well below these boundaries (see the green and blue lines). This behavior is easy to understand - the collinear radiation leads to nonzero jet masses, or equivalently, to a $d\log d\Gamma/dz$ squark decay distribution with most of the events having $z<1$ (two-body decays have $z=1$), see Figure \ref{fig:DW}. This in turn means that the jet energy is smaller than in the two body decay, cf. \eqref{pchi}, softening the $E_1+E_2$ spectrum. The effect is present, but less pronounced, also at the upper edge of the $E_1+E_2$ distribution. The original {\tt PYTHIA} distributions (before reweighting) are shown with gray lines.

The extraction of $M$, and $M_\chi$ from the $E_1+E_2$ distribution is still possible, as indicated by the fact that the $E_1+E_2$ distributions shifts significantly between the $M_\chi=0.5\TeV$ and $M_\chi=1\TeV$ benchmarks. However, one would need to use the full matrix element and not just the edges, in this way controlling the shift of the edges due to the soft and collinear radiations. In addition to the NLL + NLO and NNLL+NLO decay width distributions that we have calculated in the present manuscript, one would also control other systematics and theoretical uncertainties. 
The method, for instance, requires precise knowledge of the center of mass energy, which can be potentially distorted by beamstrahlung~\cite{Datta:2005gm, Schulte:1999xb} and initial state radiations (ISR), causing sizable uncertainties in the measurements of the edges, see, e.g., Ref.~\cite{Frank_Simon} .

\begin{figure}[!]
\begin{center}
\begin{tabular}{cc}
\hspace{-10pt} \includegraphics[scale=0.4]{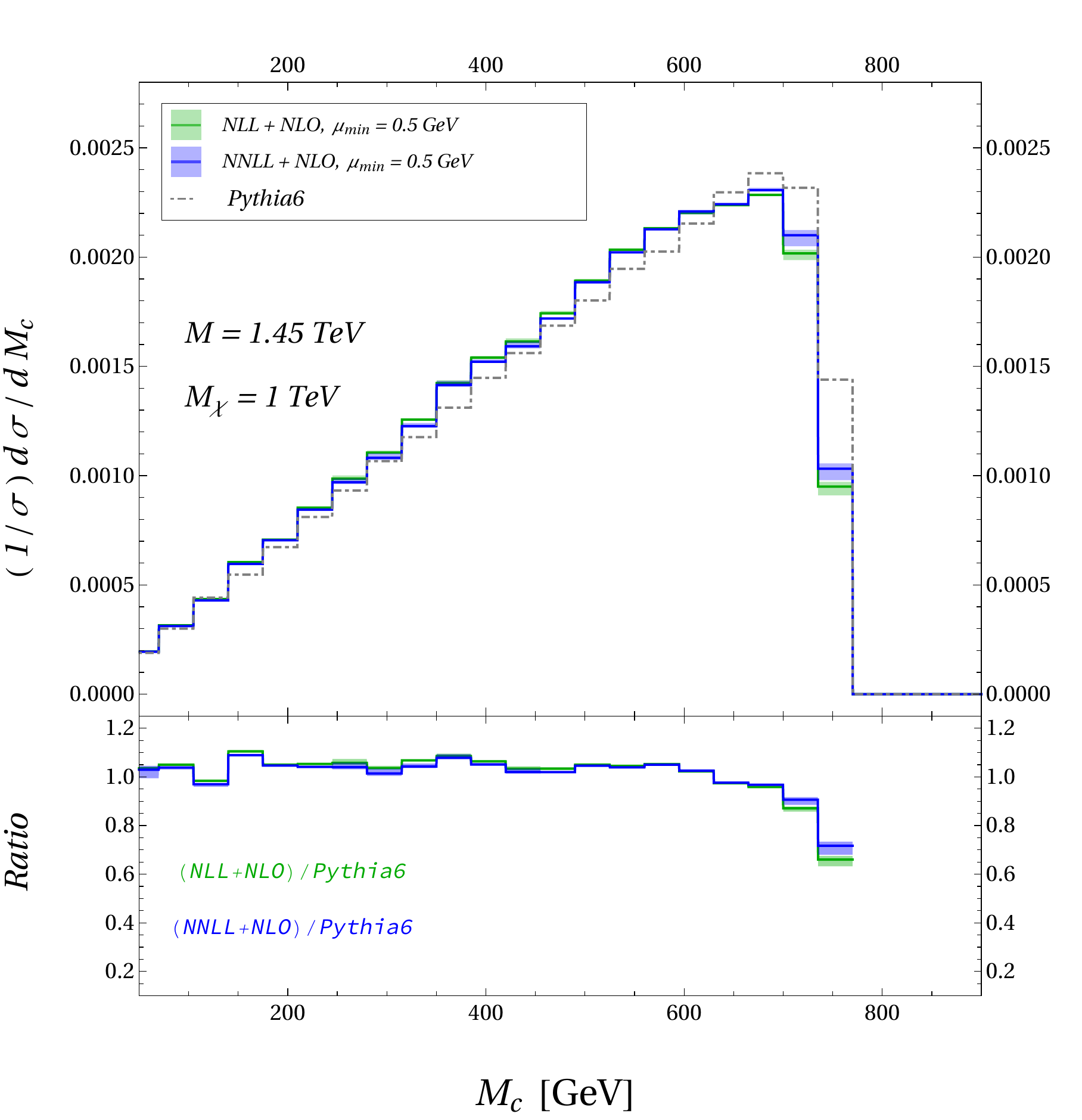} &
\hspace{+10pt} \includegraphics[scale=0.4]{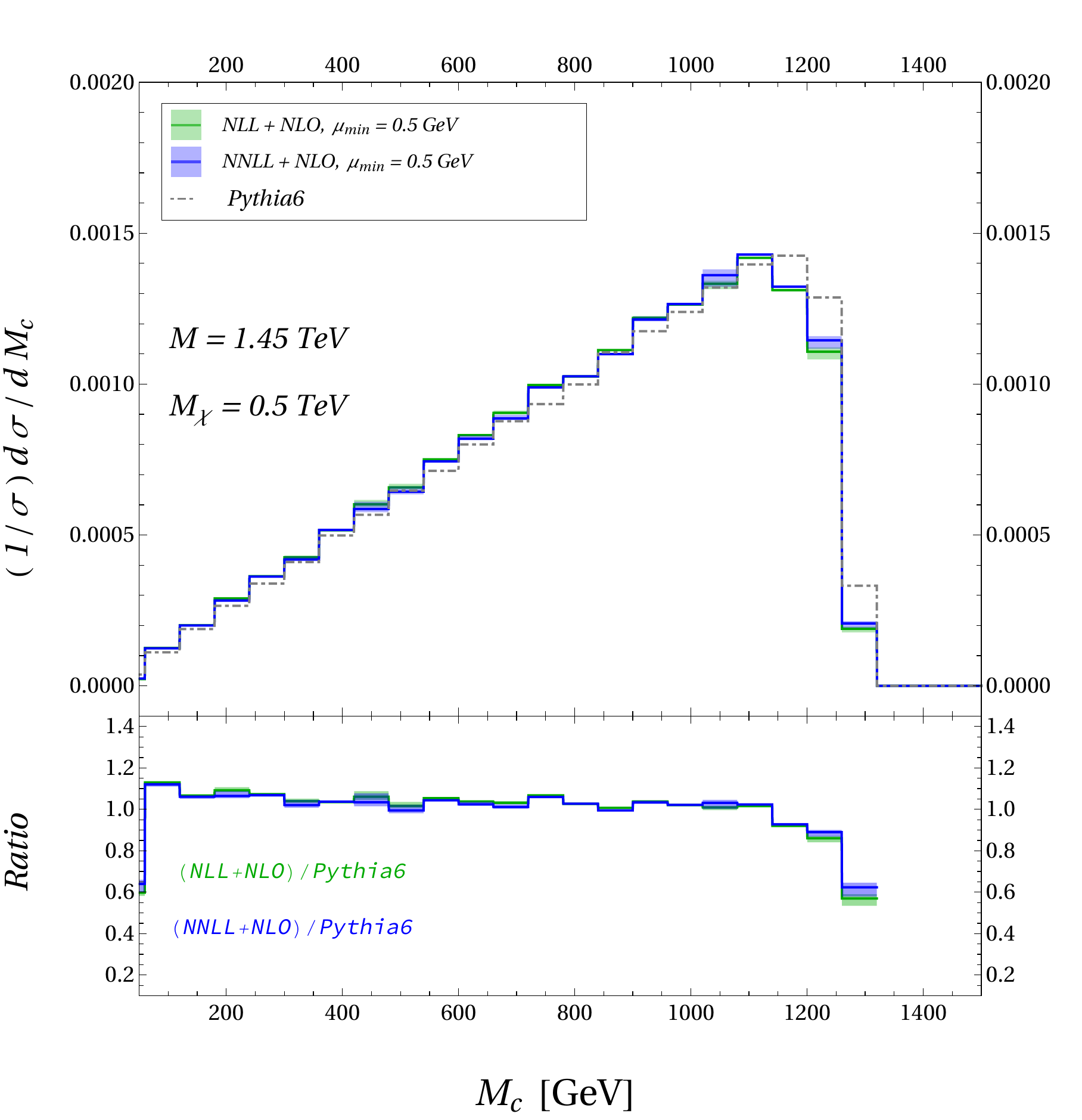} 
\vspace{-0.5cm}
\end{tabular}
\end{center}
\caption{\label{fig:MCPlots} \baselineskip 3.0ex 
The $M_C$ distributions for (left) $M_{\chi}= 1 \TeV$ and (right) $M_{\chi}= 500 \GeV$ with fixed  $M = 1.45 \TeV$.
} 
\end{figure}

An alternative mass measurement method exploits the kinematic variable $M_C$, invariant under contra-linear boosts of equal magnitude,
\begin{eqnarray} 
M_C &=& \sqrt{ \big( E_{q, 1} + E_{q, 2} \big)^2 -  \big( \vec{p}_{q, 1} -  \vec{p}_{q, 2} \big)^2 }  \; .
\label{Eq:MC}
\end{eqnarray}
Here $E_{q, 1}$, $\vec{p}_{q, 1}$ and $E_{q, 2}$, $\vec{p}_{q, 2}$ are the energies and three-momenta of the two final state quarks, respectively. The maximal value of $M_C$ is reached when the two jets are co-linear. It is given by
\begin{eqnarray} \label{eq:MCmax}
M^{\rm max}_C &=& \frac{M^2 - M^2_{\chi}}{M} \;,
\label{Eq:MC_max}
\end{eqnarray}
showing that $M_C$ is sensitive to both $M$ and $M_\chi$.
The virtue of the $M_C$ variable is that it does not depend on the center of mass energy, and is therefore less susceptible to beamstrahlung distortions~\cite{Frank_Simon}. 
Similarly to the $E_1+E_2$ distribution, the collinear and soft radiations cause the $M_C$ spectrum to soften. However, as can be seen in Figure \ref{fig:MCPlots}, the effect is more pronounced at the maximal value of $M_C$, which is exactly the quantity that enters the determination of $M$ and $M_\chi$. Comparing the shift in the distributions for $M_\chi=0.5\TeV$ and $M_\chi=1\TeV$ one sees that the LO sensitivity to $M,M_\chi$, Eq. \eqref{eq:MCmax}, still applies to a good extent also to the resummed distribution with $M_C$ constructed using the two hardest jets. For instance, for the numerical examples in Fig. \ref{fig:MCPlots} there are still appreciable numbers of events within ${\mathcal O}(5\%)$ of $M_C^{\rm max}$, with the peak of the distribution shifted by ${\mathcal O}(10-20\%)$ at NNLL+NLO compared to Pythia. This gives a rough sense of associated errors on $M_C^{\rm max}$ due to the softening of distributions in the case of limited statistics available in an experiment. However, once CLIC collects enough statistics a precise determination of $M_C^{\rm max}(M, M_\chi)$ using a matrix element method based on resummed distributions can be attempted.

\begin{figure}[!]
\begin{center}
\begin{tabular}{cc}
\hspace{-10pt} \includegraphics[scale=0.4]{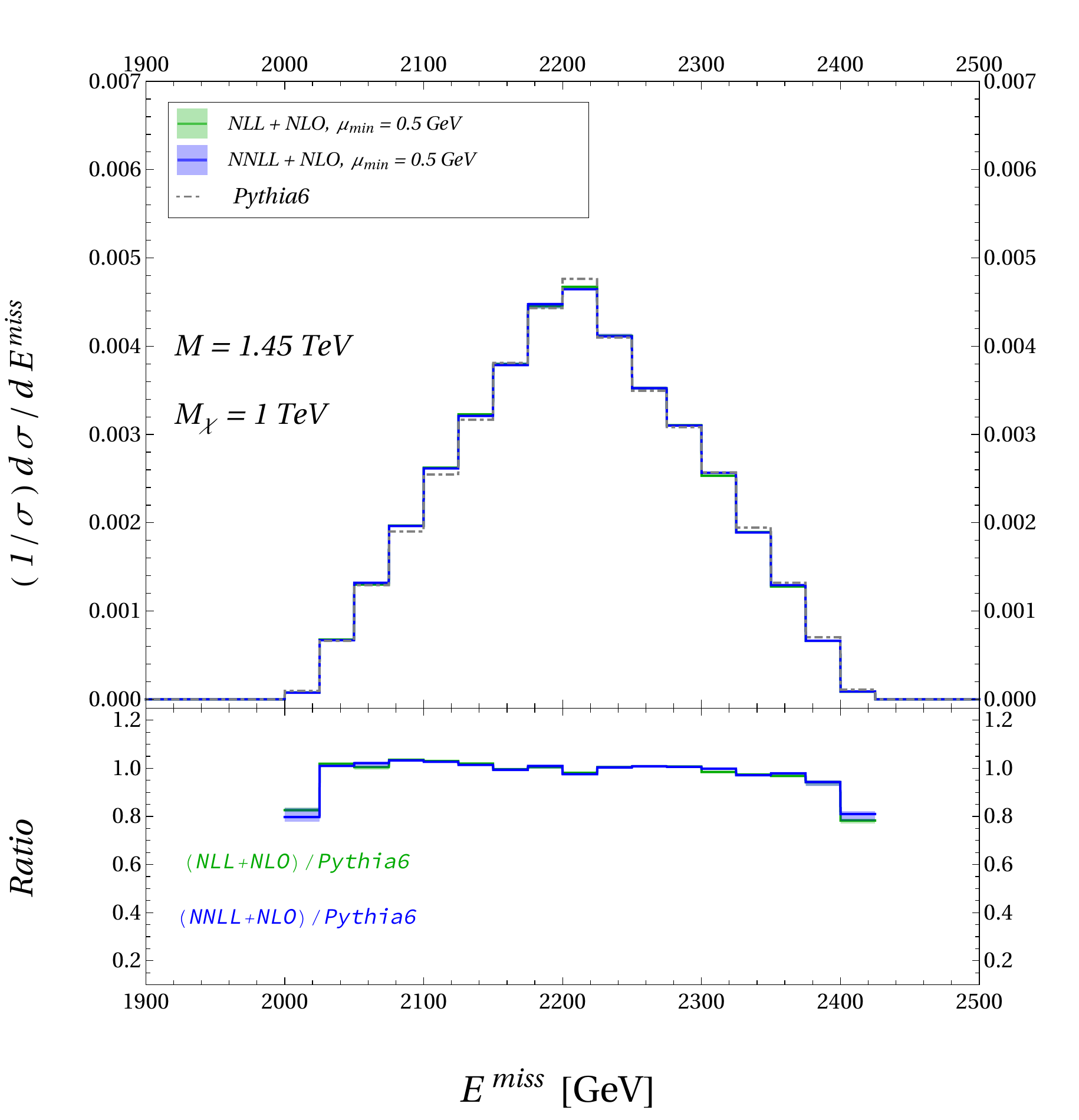} &
\hspace{+10pt} \includegraphics[scale=0.4]{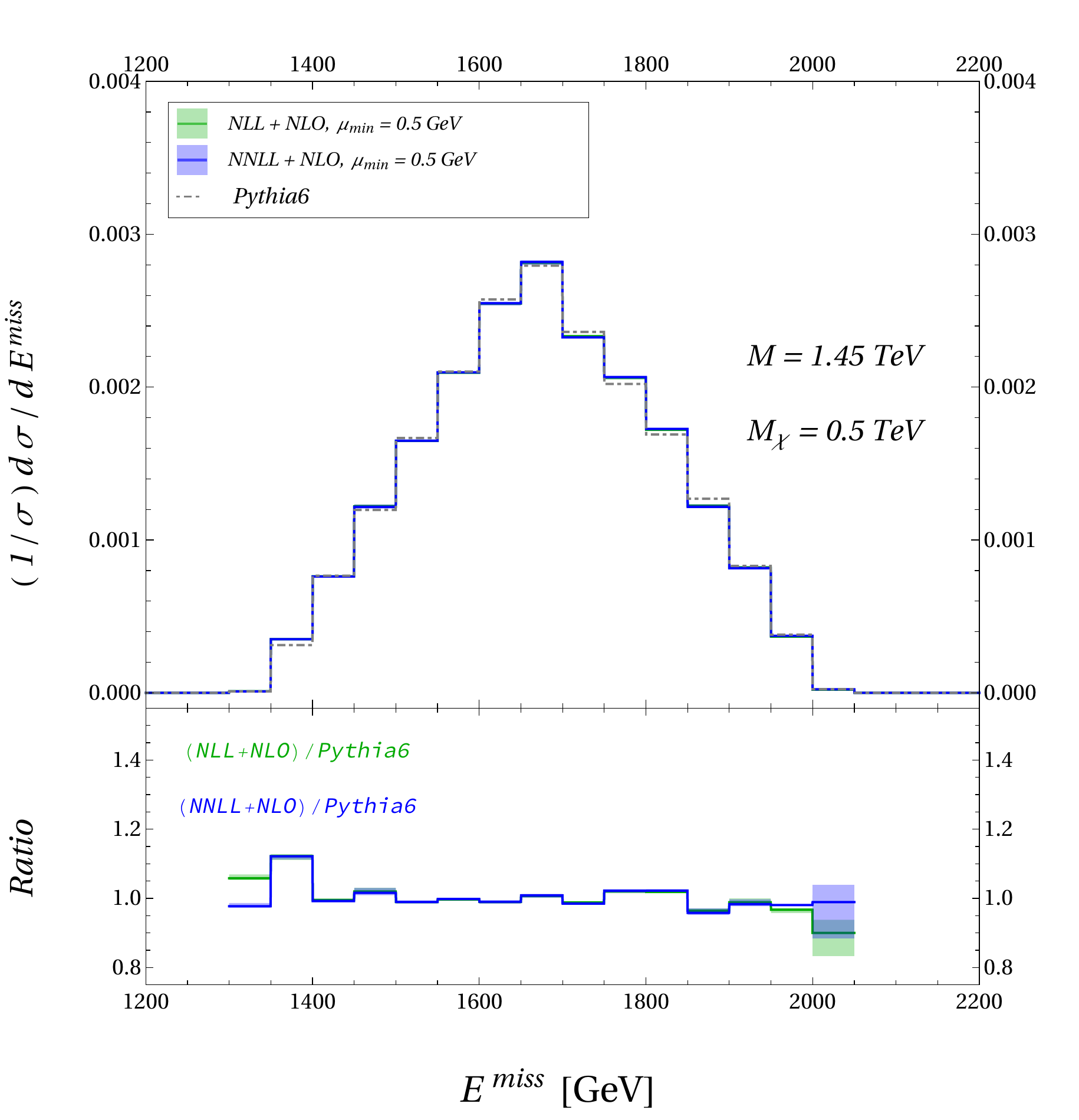}
\vspace{-0.5cm}
\end{tabular}
\end{center}
\caption{\label{fig:MissEPlots} \baselineskip 3.0ex 
The missing energy distributions for (left) $M_{\chi}= 1 \TeV$ and (right) $M_{\chi}= 500 \GeV$ with fixed  $M = 1.45 \TeV$. 
} 
\end{figure}

Finally, we show in Figure \ref{fig:MissEPlots} the {\tt PYTHIA} (gray), NLL+NLO (green), and NNLL+NLO (blue) missing energy $E^{\text{miss}}$ distributions. Here the $E^{\text{miss}}$ is due to the two neutralinos in the final state, and we do not include any detector effect. Unlike the other two observables, $E_1+E_2$ and $M_C$, the effect of resummations is negligible for the $E^{\text{miss}}$ distribution. This is because the neutralino mass is too heavy for the effect of recoiling against the soft gluon radiations from the quark-sector to be significant.

\section{Summary} \label{sec:Summary}
 
In this paper we have studied QCD corrections to the squark decay, $\tilde q\to q \chi$. The large logarithms that arise in the end-point region, $z \rightarrow 1$, were resummed using SCET up to the NNLL accuracy. Away from the end-point we computed hard gluon radiations at NLO. 
Finally, we provided an expression that smoothly interpolates between the NNLL and NLO results, giving the NNLL+NLO prediction for the total decay width and the decay distribution, $d\Gamma/dz$. The additional QCD radiation in the decay softens the decay distributions for many observables.  As a case study for the phenomenological impact of higher order QCD corrections we explored the methods for simultaneous measurements of squark and neutralino masses at a linear $e^{+}e^{-}$ collider based on $\sqrt{s} = 3$ TeV CLIC. 
A majority of mass measurement techniques are based on edges in kinematic distributions. Such kinematic edges are modified by having additional QCD radiation in the event. For instance, the distribution of the combined energy of the hardest two jets, $E_1+E_2$, now extends below the lower boundary that is otherwise obtained in the case of two body decays. Similarly, the distributions in the $M_C$ variable get softened near its maximal value, which is precisely the region used for the quark and neutralino mass extractions. With limited available statistics in experiments this softening of the distributions would result in a shift in measured squark and neutralino masses. The induced shift in the masses could be estimated from a matrix element based method using the NNLL+NLO resummed decay distributions that we provided. In a quantitative analysis one would also need to include additional effects such as the beamstrahlung, initial state radiation, and detector effects.

\acknowledgments 
JHK is grateful to Kyongchul Kong and Ian M. Lewis for useful discussions and suggestions during the course of this project. CK is supported by Basic Science Research Program through the National Research Foundation of Korea (NRF) funded by the Ministry of Science and ICT (Grant No.~NRF2017R1A2B4010511). JHK was supported in part by United States Department of Energy grant number DE-SC0017988 and by the University of Kansas General Research Fund allocation 2302091. JHK acknowledges the support in part by National Science Foundation under grant PHY1820860 and PHY1230860. SL was supported by Basic Science Research Program through the National Research Foundation of Korea(NRF) funded by the Ministry of Education (NRF-2018R1D1A1B07049148), by the Korea government (MEST) (NRF-2015R1A2A1A15052408). JZ acknowledges support in part by the DOE grant de-sc0011784.


\appendix

\section{Details about $\Delta$-distributions}
\label{AB}

Consider function $g(u,\e)$ that is singular at $\e=0$ with $u=0$, and define 
\begin{equation} 
A(\e)=\int^1_0 d u\, g(u,\e),  
\end{equation} 
where $A(\e)$ can have $(1/\e)^n$ poles. The $g(u,\e)$ function can be rewritten in terms of delta function and plus distributions as, 
\begin{equation}
\label{uexpan} 
g(u,\e) = A(\e) \delta (u) + [g(u,\e)]_+, 
\end{equation} 
where the plus distribution is defined as 
\begin{equation} 
\label{uplus}
\int^1_0 du \bigl[g(u)\bigr]_+ f(u) = \int^1_0 du~g(u) \Bigl[f(u)-f(0)\Bigr].
\end{equation}

For application to our results it is useful to make a change of variables,  $u=h(z)$, so that the singular point, $u=0$, is now at $z=1$, while $z\in [a,1]$, i.e., $h(1)=0$ and $h(a)=1$. When extracting the IR poles from the integral over $z$ one needs to carefully keep track all the factors due to a change of variables. Consider now the integral 
\begin{equation} 
\int^1_a dz f(z) g(u,\e)
= \int^1_0 du \Bigl(-\frac{dz}{du}\Bigr) \tilde{f}(u) \Bigl(A(\e) \delta(u)+ \big[g(u,\e)\bigr]_+\Bigr),
\end{equation}
where on the lhs $g(u,\e)=g(h(z),\e)$, while on the rhs $\tilde{f}(u) =f(h^{-1}(u))=f(z)$.
Using Eq.~(\ref{uplus}) we can rewrite the above expression as 
\begin{equation}
\begin{split} 
\label{Del}
\int^1_a dz f(z) g(u,\e) &=
\int^1_0 du \Bigl(-\frac{dz}{du}\Bigr) \tilde{f}(u) A(\e) \delta(u)
-\int^1_0 du g(u,\e) \Bigl(\frac{dz}{du}\tilde{f}(u) - \frac{dz}{du}\Bigr|_{u=0} \tilde{f} (0) \Bigr)  
\\
&=\int^1_a dz f(z) A(\e)  \delta(1-z) \Big/ \Bigl|\frac{du}{dz}\Bigr|_{z=1}
\\
&\quad+\int^1_a dz\, g(u,\e) \Bigl[f(z) - f(1) \frac{du}{dz}\Big/\Big(\frac{du}{dz}\Bigr|_{z=1}\Big) \Bigr],
\end{split}
\end{equation}
where in the last two lines $u$ should be viewed as a function of $z$, $u=h(z)$.
This means that in terms of the $z$-space distributions we have 
\begin{equation} 
\label{Del1}
g(u,\e) = A(\e) \Bigl|\frac{du}{dz}\Bigr|^{-1}_{z=1} \delta(1-z) + \Bigl[g(u,\e)\Bigr]_{\Delta},
\end{equation}
where we have defined $\Delta$-distribution as 
\begin{equation} 
\label{Del2} 
\int^1_a dz f(z) \Bigl[g(u,\e)\Bigr]_{\Delta} = 
\int^1_a dz \, g(u,\e) \Biggl[f(z) - f(1)\frac{du}{dz}\biggr/ \Big(\frac{du}{dz}\Bigr|_{z=1}\Big)\Biggr].
\end{equation}
Note that term in the bracket multiplying $g(u,\epsilon)$ in the right side tends to zero as $z\to 1$. 

In obtaining Eq.~(\ref{Wonel}) in the main text, we used the change of variables,  $u=(1-z)/(1-x/z)$, for which $du/dz = -(1-2x/z+x/z^2)/(1-x/z)^2$, and $du/dz|_{z=1}=-1/(1-x)$. A function $g(u,\e) = 1/u^{1+\e}$, can then be expressed in terms of  $z$-space distributions as 
\begin{equation} 
\frac{1}{u^{1+\e}} = -\frac{1-x}{\e}\delta(1-z) + \Bigl[\frac{(1-x/z)^{1+\e}}{(1-z)^{1+\e}}\Bigr]_{\Delta},
\end{equation}
The $\Delta$-distribution can be further expanded by $\epsilon$, 
\begin{equation} 
\int^1_{\sqrt{x}} dz \Bigl[\frac{(1-x/z)^{1+\e}}{(1-z)^{1+\e}}\Bigr]_{\Delta} f(z) 
= \int^1_{\sqrt{x}} dz \frac{1-x/z}{1-z} \Bigl[ f(z) - \frac{(1-2x/z+x/z^2)(1-x)}{(1-x/z)^2} f(1) \Bigr]+\mO(\e).
\end{equation}

In calculating the Feynman diagrams (c) and (d) in Fig.~\ref{fig2}, one encounters the following integral
\begin{equation} 
\int^1_0 d\alpha\, \frac{\alpha^{-\e}(1-\alpha)^{-\e}}{(1-u)\alpha +u }
= \frac{\big(\Gamma(1-\e)\big)^2}{u} ~{}_2\tilde{F}_1 \big(1, 1-\e; 2-2\e; \tfrac{u-1}{u}\big),
\end{equation}  
which is divergent for $u=0$ and $\e = 0$.
For extraction of $1/\e$ poles we can use that
\begin{equation} 
\label{intF}
\int^1_0 du\, u^{-2-\e} {}_2\tilde{F}_1 \big(1, 1-\e; 2-2\e; \tfrac{u-1}{u}\big) = - \frac{1}{4\e^3 \Gamma(-2\e)},
\end{equation}
where the regularized hypergeometric function ${}_2\tilde{F}_1$  simplifies at $\e =0$ to 
\begin{equation} 
{}_2\tilde{F}_1 \big(1, 1; 2; \tfrac{u-1}{u}\big) = - \frac{u\, \ln u}{1-u}.
\end{equation}
Applying Eq.~(\ref{Del1}) then leads to
\begin{equation}
\begin{split}
u^{-2-\e} {}_2\tilde{F}_1 \big(1, 1-\e; 2-2\e; \tfrac{u-1}{u}\big) =& 
-\frac{1-x}{4\e^3 \Gamma(-2\e)} \delta(1-z) 
\\
~~~~~~~~~~&-\left[\frac{(1-x/z)^2}{(1-z)(z-x/z)} \ln\frac{(1-z)}{(1-x/z)}\right]_{\Delta}
+\mO(\e).
\end{split}
\end{equation}

\section{Two-loop jet and soft functions}
\label{two_loop}

In this appendix we give the log enhanced two-loop contributions in jet and soft functions, Eqs. \eqref{JJ}, \eqref{SS}.
First, we consider the moments of the jet function
\begin{align}
\tilde{J} [\bar{N}] = \int^{1}_{0} dz~ z^{-1+N} J(Q^2 (1-z)) \;.
\end{align}
In the large N limit we can write
\beq
\begin{split}
\label{J_moment}
\tilde{J} [\bar{N}] =  1+&\frac{\alpha_s}{4\pi} \Bigl( j^{(1)}_0 +  j^{(1)}_1 L_j + j^{(1)}_2 L^2_j      \Bigl) \\
 + &\Bigl( \frac{\alpha_s}{4\pi}  \Bigl)^2 \Bigl(  j^{(2)}_1 L_j + j^{(2)}_2 L^2_j  + j^{(2)}_3 L^3_j  + j^{(2)}_4 L^4_j      \Bigl) \;,
 \end{split}
\eeq
where $L_j = \ln ( \mu^2 \bar{N}/ (M^2(1-\rc)) )$ and $\bar{N} = N \exp(\gamma_E)$. Note that  in the second line of the equation we dropped the non-logarithmic term, $\big( \alpha_s/(4\pi)  \big)^2j^{(2)}_0$. The RG equation becomes
\beq
\begin{split}
\frac{d \tilde{J}  }{d \ln \mu} =  \gamma_J \tilde{J} = & \frac{\alpha_s}{4\pi}  \Bigl( \hat{\gamma}_{J,0} + 2 \Gamma_0 L_j \Big) + \Bigl( \frac{\alpha_s}{4\pi}  \Bigl)^2
 \Bigl[  \hat{\gamma}_{J,1} +  \hat{\gamma}_{J,0} j^{(1)}_0 + L_j \big(  \hat{\gamma}_{J,0} j^{(1)}_1 + 2 \Gamma_0 j^{(1)}_0 + 2 \Gamma_1 \big) \\
 &+ L^2_j  \big(  \hat{\gamma}_{J,0} j^{(1)}_2 + 2 \Gamma_0 j^{(1)}_1 \big) +  2 \Gamma_0 j^{(1)}_2 L^3_j    \Bigl] \;,
 \end{split}
\eeq
where we have use that $ \gamma_J = 2 \Gamma_C L_j +\hat{\gamma}_J$, with $\hat{\gamma}_{J}=\sum_{k=0} \hat{\gamma}_{J,k} (\alpha_s/(4\pi))^k $. 
From this we can obtain the coefficients of the $\alpha_s^2$ log terms in Eq.~(\ref{J_moment}),
\begin{align}
 j^{(2)}_1 &=  \frac{1}{2} \Bigl( \hat{\gamma}_{J,1} + \hat{\gamma}_{J,0} j^{(1)}_0 + 2 \beta_0 j^{(1)}_0 \Bigl), \\ 
 j^{(2)}_2 &=  \frac{1}{4} \Bigl(  \big( \hat{\gamma}_{J,0} + 2 \beta_0 \big) j^{(1)}_1 + 2 \Gamma_0 j^{(1)}_0 + 2 \Gamma_1 \Bigl), \\ 
 j^{(2)}_3 &=  \frac{1}{6} \Bigl(  \big( \hat{\gamma}_{J,0} + 2 \beta_0 \big) j^{(1)}_2 + 2 \Gamma_0 j^{(1)}_1  \Bigl),  \\ 
  j^{(2)}_4 &=  \frac{1}{8} \Bigl( 2 \Gamma_0 j^{(1)}_2 \Bigl) \;.
\end{align}
In a similar way, we write for the soft function
\beq
\begin{split}
\label{S_moment}
\tilde{S} [\bar{N}] =  1+&\frac{\alpha_s}{4\pi} \Bigl( s^{(1)}_0 +  s^{(1)}_1 L_s + s^{(1)}_2 L^2_s      \Bigl) \\
+& \Bigl( \frac{\alpha_s}{4\pi}  \Bigl)^2 \Bigl(  s^{(2)}_1 L_s + s^{(2)}_2 L^2_s  + s^{(2)}_3 L^3_s  + s^{(2)}_4 L^4_s      \Bigl) \;,
\end{split}
\eeq
where $L_s = \ln ( \mu^2 \bar{N}/ M^2 )$, and the coefficients of the $\alpha_s^2$ log terms are determined to be 
\begin{align}
 s^{(2)}_1 &=  \frac{1}{2} \Bigl( \hat{\gamma}_{s,1} + (\hat{\gamma}_{s,0}  + 2 \beta_0 ) s^{(1)}_0 \Bigl), \\ 
 s^{(2)}_2 &=  \frac{1}{4} \Bigl(  \big( \hat{\gamma}_{s,0} + 2 \beta_0 \big) s^{(1)}_1 + A_s ( \Gamma_0 s^{(1)}_0 + \Gamma_1 )  \Bigl),  \\ 
 s^{(2)}_3 &=  \frac{1}{6} \Bigl(  \big( \hat{\gamma}_{s,0} + 2 \beta_0 \big) s^{(1)}_2 + A_s \Gamma_0 s^{(1)}_1  \Bigl),  \\
  s^{(2)}_4 &=  \frac{1}{8} \Bigl( A_s \Gamma_0 s^{(1)}_2 \Bigl),
\end{align}
where $A_s = -1$, $\gamma_S = A_s \Gamma_C L_s +\hat{\gamma}_S$, and $\hat{\gamma}_{S}=\sum_{k=0} \hat{\gamma}_{S,k} (\alpha_s/(4\pi))^k$.


\bibliography{lit}

\end{document}